\documentclass[final,twocolumn,showpacs,preprintnumbers,amsmath,amssymb,prc,nofootinbib]{revtex4-1}
\usepackage{graphicx}% Include figure files
\usepackage{dcolumn}% Align table columns on decimal point
\usepackage{bm}% bold math
\usepackage{multirow}
\usepackage{color}
\usepackage{amsmath}
\usepackage{mathtools}
\usepackage{textcomp}
\usepackage{amssymb}
\usepackage{url}

\usepackage{breakurl}

\usepackage{longtable}
\setlongtables

\definecolor{Darkgreen}{rgb}{0,0.4,0}
\usepackage[breaklinks=true,linkbordercolor={1 1 1}]{hyperref}
\hypersetup{
colorlinks = true,    % set true if you want colored links
linktoc    = all,     % set to all if you want both sections and subsections linked
linkcolor  = blue,    % choose some color if you want links to stand out
citecolor  = Darkgreen
}

\usepackage[caption=false]{subfig}
\usepackage{dcolumn}
\usepackage{natbib}
\usepackage{color}
\usepackage[normalem]{ulem}
\usepackage{soul,xcolor}
\usepackage{xfrac}
\usepackage{cancel}

\usepackage{paralist}

\usepackage{mathdesign}
\usepackage{booktabs}

\newcommand{\nuc}[2]{$^{{#1}}${#2}}

\newcommand{\scikitlearn}{\texttt{scikit-learn}}

\usepackage{draftwatermark}
\SetWatermarkText{\textsc{Draft}}
\SetWatermarkScale{6}

\def\etal{\emph{et~al.}\ }

\newcommand{\be}{\begin{equation}}
\newcommand{\ee}{\end{equation}}

\newcommand{\bea}{\begin{eqnarray}}
\newcommand{\eea}{\end{eqnarray}}

\begin{document}

\title{A novel Machine-Learning method for spin classification of neutron resonances}

\author{G. P. A. Nobre}
\email[Corresponding author: ]{gnobre@bnl.gov}
\affiliation{National Nuclear Data Center, Brookhaven National Laboratory, Upton, NY 11973-5000, USA}

\author{D. A. Brown}
\affiliation{National Nuclear Data Center, Brookhaven National Laboratory, Upton, NY 11973-5000, USA}

\author{S. J. Hollick}
\affiliation{Department of Physics, Yale University, New Haven, CT, 06520, USA}

\author{S. Scoville}
\affiliation{University of Pittsburgh, Pittsburgh, PA 15217, USA}
\affiliation{Rensselaer Polytechnic Institute, Troy, NY, 12180, USA}

\author{P. Rodr\'{i}guez}
\affiliation{Pacific Northwest National Laboratory, Richland, WA 99354, USA}
\affiliation{University of Puerto Rico, Mayag\"uez Campus, Mayagüez, 00682, Puerto Rico}

%\author{M. Fucci}
%\affiliation{University At Albany, Department of Physics}

%\author{S. Ruiz}
%\affiliation{Georgia Institute of Technology, School of Physics}

%\author{R.-M. Crawford}
%\affiliation{Willamette University, Department of Physics}

%\author{A. Coles}
%\affiliation{National Nuclear Data Center, Brookhaven National Laboratory, Upton, NY 11973-5000, USA}

%\author{M. Vorabbi}
%\affiliation{National Nuclear Data Center, Brookhaven National Laboratory, Upton, NY 11973-5000, USA}

%===================================================================================================================
\begin{abstract}

The performance of nuclear reactors and other nuclear systems depends on a precise understanding of the neutron interaction cross sections for materials used in these systems. These cross sections exhibit resonant structure whose shape is determined in part by the angular momentum quantum numbers of the resonances.
The correct assignment of the quantum numbers of neutron resonances is, therefore, paramount.
In this project, we apply machine learning to automate the quantum number assignments using only the resonances' energies and widths and not relying on detailed transmission or capture measurements.
The classifier used for quantum number assignment is trained using stochastically generated resonance sequences whose distributions mimic those of real data.
We explore the use of several physics-motivated features for training our classifier.
These features amount to out-of-distribution tests of a given resonance's widths and resonance-pair spacings.
We pay special attention to situations where either capture widths cannot be trusted for classification purposes or where there is insufficient information to classify resonances by the total spin $J$.
We demonstrate the efficacy of our classification approach using simulated and actual $^{52}$Cr resonance data.

\end{abstract}
%\pacs{\textcolor{red}{Look up pacs codes!!}}
\date{\today}

\maketitle

\setstcolor{red}

%===================================================================================================================
\section{Introduction}
\label{sec:intro}

Neutron scattering and reaction data for neutron energies ranging from 10$^{-5}$ eV to 20 MeV are needed for simulations of nuclear systems in nuclear fission and fusion energy production, stockpile stewardship, non-proliferation, etc.~\cite{ENDF-VIII.0}.
For energies below that typical of fission neutrons, $\sim 1$ MeV, normally only elastic and capture (and fission for actinides) channels are open.
For all but the lightest nuclei, these reaction channels all exhibit strong resonant structure that we identify with the energy levels of the compound nucleus formed by the capture of the neutron into the target state~\cite{Satchler:1980}.

The double differential capture or elastic scattering cross sections are completely determined by the set of resonance energies, the decay widths to each of the observed reaction channels and incident neutron orbital angular momentum $L$ and the total angular momentum $J$ characterizing these reaction channels \footnote{In the $JLS$ coupling scheme which we use in this work, there is also the total spin of the incident channel, $S$.  This can usually be determined from knowledge of $L$ and $J$ for s- and p-wave resonances.}, when described using R-matrix theory~\cite{Lane:1958,Blatt:1952}.
We cannot predict the energies and widths of the resonances in any nuclei other than the lightest systems with current theoretical and computational approaches.
The resonance energies and widths must be determined by fitting experimental transmission or cross section measurements.
To complicate matters, the shape of the R-matrix fitting function is heavily dependent on the quantum numbers ($L, J$) assigned to the particular resonance.

Codes, such as SAMMY~\cite{SAMMY} and REFIT~\cite{REFIT}, use a Generalized Least-Squares Fitting routine derived from a linearized version of Bayes' Equation.
Conventional evaluations based on SAMMY or REFIT  require significant preparation by an evaluator to establish reliable prior estimates of the widths, energies and $(L, J)$ quantum numbers of the resonances, ensuring that one is sufficiently close to the $\chi^2$ minimum for the fit to be well founded.
Unfortunately, the shear number of known resonances in a typical evaluation makes this endeavor tedious and time-consuming.
Furthermore, this step of the evaluation is subjective, relying on the experience of the evaluator and, therefore, it is hardly reproducible.
This fact leads to significant amounts of unquantified uncertainty in the final evaluation.

\begin{figure*}
\begin{center}
\includegraphics[width=0.6\textwidth]{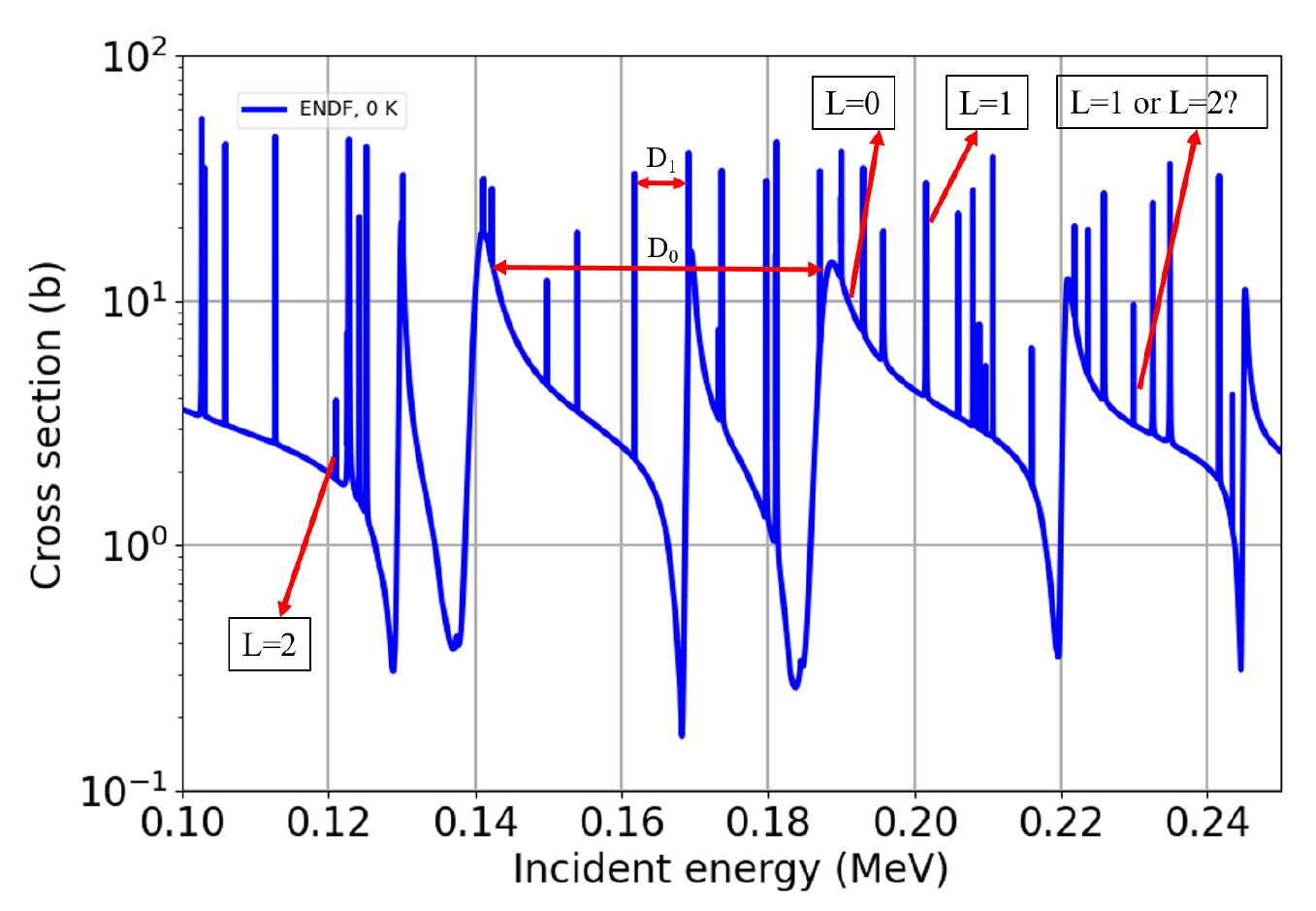}
\end{center}
\caption{
\label{fig:classificationbyhand}
A portion of a typical resonance region cross section; namely, the elastic cross section for $^{238}$U extracted from the ENDF/B-VIII.0 evaluated file~\cite{ENDF-VIII.0}, as a illustrative representation of resonance properties.
In this figure we show several $L=0,1,$ and 2 resonances and label the spacing $D_L$ between a pair of $L=0$ and $L=1$ resonances, respectively.
We note that it is often quite difficult to discern between a $L=1$ and $L=2$ resonance.
Determining the $J$ quantum number is significantly more challenging as indicated in the main text.  
}
\end{figure*}

There are a number of experimental techniques that can help determine the incident neutron orbital angular momentum $L$ and the total angular momentum $J$ of each resonance including study of the low-energy $\gamma$-ray cascades from neutron capture events detected by Ge-Li detectors, $\gamma$-ray multiplicity methods, and measurements with polarized neutron beams and polarized targets.  In the best case, angular distribution data for scattered neutrons or emitted photons are available and can be used to determine the $L$ and $J$ of the resonance.
Between the hundreds, or thousands, of resonances per nuclide in an evaluation and the technical complexity of some of these techniques, they are often not used in practice.
Fig.~\ref{fig:classificationbyhand} shows a representation of measured cross-section data where two distinct resonance shapes are observable: a wide and asymmetric shape corresponding to s-wave ($L=0$) resonances and a narrow and symmetric resonance shape corresponding to p-wave ($L=1$) resonances.
Note, however, that the visible distinction in the experimental data between the two shapes diminishes with increasing incident neutron energy.

The current practice world-wide is for the resonance evaluator to visually inspect the experimental cross-section, yield or transmission data, such as in Fig.~\ref{fig:classificationbyhand}, sometimes for thousands of resonances, and make the spin assignments for each resonance.
As mentioned before, this part of the process is 
\begin{inparaenum}[i)]
\item very time consuming for the evaluator,
\item not fully reproducible,
\item does not result in uncertainty estimate on the correct resonance spin assignment, and 
\item has significant impact on the angular distributions and, therefore, on the modeling of neutron transport in nuclear systems.
\end{inparaenum}
Furthermore, visual inspection of the resonance shape in  experimental cross-section data can only determine the orbital angular momentum $L$ (s-wave, p-wave) corresponding to each resonance and not the total angular momentum $J$.
The evaluator is left to chose the total angular momentum by observing small changes in the interference pattern between resonances of the same orbital angular momenta.

Moving beyond a pure experimental approach, there are some early attempts at information-theoretic techniques for resonance spin classification.
Ref.~\cite{Georgopulos:1981} was the first to suggest using random matrix theory (RMT) inspired metrics to determine the fraction of missing levels using stochastically generated resonances.
Ref.~\cite[p. 81]{Froehner:2000}  suggests probabilistic assignment based on consideration of width distribution.
This concept is expanded on by Mitchell \etal~\cite{Mitchell:2004}.
The series of papers by Mulhall \etal examine the use of $\Delta_3$ statistic to infer the purity of a spin sequence~\cite{Mulhall:2011,Mulhall:2010,Mulhall:2009,Mulhall:2007}.
Finally, there is a pair of reports by Mitchell and Shriner estimating the fraction of missing or misclassified resonances~\cite{Mitchell:2009,Shriner:2011} using various RMT-inspired metrics.

In this study, we aim to develop a more reliable, automated and reproducible method through the utilization of a variety of standard machine-learning classification algorithms.
The classifiers used in this study can be found in the \scikitlearn\ python module~\cite{scikit-learn}. In the recent years many statistical and computational tools aiming to mimic the way the human brain functions to identify patterns and learn how to solve problems, broadly named Machine-Learning (ML) methods, have been optimized and packaged for general purpose. These have been applied to an extremely wide variety of applications. Our goal is to leverage such methods and foundation to develop a new and reproducible approach to the classification of neutron resonances.

This paper is organized as follows.  In Section \ref{sec:properties}, we review the relevant statistical and average properties of neutron resonances.
Using these properties, we develop a set of machine-learning features allowing us to recast the quantum number assignment problem as a machine-learning problem in Section \ref{sec:recast}.  In Section \ref{sec:application}, we apply our machine-learning approach to $n+^{52}$Cr neutron resonances.
In Section  \ref{sec:conclusion}, we provide a summary and outlook. As a reference, we present in Appendix~\ref{app:ML_Glossary} the definition of ML terms and concepts used throughout the text.

%===================================================================================================================
\section{Statistical properties of resonances}
\label{sec:properties}

Although the experimental situation is complicated, there are results from both nuclear reaction and RMTs that  will make our classification problem more tractable.
Here we do not aim for a review of theory of neutron resonances as there are many other sources for that (e.g., Refs.~\cite{Atlas,Froehner:2000}).
Rather, we highlight results that impact our resonance classification task.

% --------------------------
\subsection{$JLS$ coupling}
% --------------------------

Our classification scheme focuses on the $L$ and $J$ quantum numbers.
As R-matrix analysis of neutron resonances is nearly always done using the $JLS$ coupling scheme~\cite{Lane:1958}, it is, therefore, useful to expand on it.
The $JLS$ scheme describes the coupling of the incident neutron with orbital angular momentum $L$, spin $1/2$, and target nucleus spin $I_t$ to the total angular momentum $J$.

In the $JLS$ coupling scheme, the two particles participating in a reaction channel have their spins coupled to a total channel spin $S$.
For an entrance channel with target nucleus spin and parity $I_t^\Pi$ and incident neutron with spin and parity $I_n=\frac{1}{2}^+$, the total channel spin may take values $S=|I_t-I_n|, \dots, I_t+I_n$.
Since neutrons have spin $1/2$, this limits $S$ to at most two values, $S=|I_t-1/2|$ and $I_t+1/2$.
For a spin zero nucleus, only $S=1/2$ is allowed.

The total channel spin $J$ then may take values $J=|L-S|, \dots, L+S$.
For a spin zero nucleus, $J$ is limited to $1/2$ for s-wave resonances ($L=0$).
For $L>0$, $J$ takes two values $L-1/2$ and $L+1/2$.
For higher spin target nuclei, $J$ can take many values.

Additional consideration of the parity of the neutron and target limits the potential values of $J$ somewhat but does not change the essential problem that there are usually many possible values of $J$ for a given $L$.
Fr\"ohner provides a table of allowed values in Ref.~\cite[ Table 2, page 50]{Froehner:2000}.

In any event, these considerations of angular momentum limit the possible labels we can assign to a resonance sequence to a tractable number.
In some cases, these considerations completely determine the $J$ value for a given $L$, at least in the case of nuclei with a spin zero ground state.

% -------------------------------
\subsection{Random matrix theory}
% -------------------------------
\label{subsec:rmt}

Within a sequence of resonances with the same $L$ and $J$ (and perhaps $S$), which defines a spingroup, the question arises as to whether there are qualities of the resonances and/or the entire sequence that can inform the classification task. 
The answer is affirmative if one considers the direct results of  RMT.

In random matrix theory, we make a bold and somewhat surprising assumption about the compound nuclear states and their couplings to the outside space:
we assume that the Hamiltonian governing the system's couplings between states obeys all relevant symmetries (so it is invariant under an orthogonal transformation) but is otherwise made of random numbers drawn from a normal distribution.
The collection of all such Hamiltonians with a given dimension and coupling scale $D$ is the Gaussian Orthogonal Ensemble (GOE).
It can be shown that the eigenvalues of these GOE Hamiltonians (which we identify with compound nuclear states and hence resonance energies) have a joint probability density given by Refs.~\cite{Weidenmueller:2009,Mehta:1967}
\begin{equation}
    \begin{array}{rl}
P_{\mathrm{GOE}}(H)d[H] =& \displaystyle N_{0} dO \exp\left[-\frac{N}{4\lambda^2}\sum_k E^2_k\right]\\
  & \displaystyle \times\prod_{l<m}|E_l-E_m|\prod_{n} dE_n.
    \end{array}
    \label{eq:goepdf}
\end{equation}
Here $dO$ is the de Haar measure of the integral, $N_0$ is a normalization constant, $N$ is the dimension of the space (assumed large), $E_\mu$ are the eigenvalues of the Hamiltonian $H$, and the constant $\lambda=N*D/\pi$ with $D$ being the mean spacing between states.

By itself, Eq.~\eqref{eq:goepdf} cannot be used as a ML feature in our problem.
Even for small $N$, the probabilities of a given configuration of energies is numerically very small even if a particular configuration has a high relative probability compared to other configurations.
Thus, use of this as a feature would be plagued by numerical precision issues.

Eq.~\eqref{eq:goepdf} can be used to derive correlations between the resonance energies of spin group sequences of nearly any length.
This will allow us to develop classification features that are ``local'' in that they depend only on a resonance and its nearest neighbors in the sequence.
Thus, classification errors in the sequence far from a given resonance will not impact its own classification.
The most interesting correlations for our purposes are the short-range correlations characterized by the nearest neighbor spacing distribution (NNSD) and the spacing-spacing distribution (SSD).
Eq.~\eqref{eq:goepdf} alone does not fully motivate the last interesting set of correlations, the width distributions as we will discuss below.

\paragraph{Nearest neighbor spacing distribution (NNSD) - }
The spacing between the  $n^{\mathrm{th}}$ resonance and the  $(n+1)^{\mathrm{th}}$ resonance is $D_n=E_{n+1}-E_n$.
From Eq.~\eqref{eq:goepdf}, one can show that the distribution of $D_n$'s follows a distribution colloquially known as ``Wigner's surmise''~\cite{Mehta:1967}:
    \begin{equation}
        P_{w}(x) = \frac{\pi x}{2} \exp\left(-\frac{\pi x^2}{4}\right)
        \label{eq:wignersurmise}
    \end{equation}
Here $x=D/\overline{D}$, where $\overline{D}$ is the average spacing.
%An apocryphal story credits the name ``Wigner's surmise'' to a guess made by Eugene Wigner in response to the presentation of the Porter-Thomas distribution at a nuclear physics conference \fixme{This reference is listed as ``apocryphal story'' \cite{apocryphal-story}}.
%This guess was subsequently described in terms of the GOE joint probability distribution above.
We note several things about this distribution: it favors spacings approximately near $\overline{D}$; the fact that it approaches zero for small spacings elegantly explains level repulsion; and it does not forbid large spacings, but strongly discourages them.  In this way, Wigner's surmise prefers a ``picket fence'' like sequence of resonances within a spingroup.

We note that a spacing distribution made of resonances from many spingroups will destroy the correlations encoded in Wigner's surmise and the nearest neighbor spacing distribution will tend toward a Poisson distribution.

\paragraph{Spacing-spacing distribution (SSD) - }
A slightly longer range correlation is the spacing-spacing correlation, denoted $\rho$:
    \begin{equation}
        \rho_n = \frac{D_n D_{n+1}}{\overline{D}^2}
        \label{eq:rho}
    \end{equation}
The distribution of spacing-spacing correlations $P_{\mathrm{ssc}}(\rho)$ is not known analytically but has been mapped out numerically~\cite{Mehta:1967}.
The mean spacing-spacing correlation is known to be $\overline{\rho}=\sum_n \rho_n/N\approx-0.27$.
The implication of the average anticorrelation between spacings is that resonance spacings tend to follow a short-long-short-long pattern.

\paragraph{Channel width distributions (CWD) - }
We can imagine expanding our random Hamiltonian to include random couplings to continuum states outside of the considered space, then looking to the poles of the resulting scattering matrix~\cite{Mitchell:2010}.
This train of reasoning eventually explains the empirically known Porter-Thomas distribution of resonance widths~\cite{Atlas,Porter:1956}:
\begin{equation}
    P_{pt}(x|\nu) = {\frac {1}{2^{\nu/2}\Gamma (\nu/2)}}\;x^{\nu/2-1}e^{-x/2}.
    \label{eq:porterthomas}
\end{equation}
Here $x=\Gamma/\overline{\Gamma}$ (where $\overline{\Gamma}$ is the average width).  We may also write this in terms of the reduced width amplitudes, $x=\gamma^2/\overline{\gamma^2}$, where $\Gamma=2P_c\gamma^2$ and $P_c$ is the penetrability factor for the channel\footnote{The channel index includes the quantum numbers of the spingroup as well as the identity of the two particles which are coupling together within this channel.} $c$ in question \cite{Froehner:2000,Atlas}.
This distribution is a $\chi^2$ distribution with $\nu$ degrees of freedom where $\nu$ represents the number of channels coupled to this spingroup with matching quantum numbers.

For moderate to large $\nu \gtrsim 5$, width distributions peak at the average channel width.
For small $\nu$, width distributions are strongly peaked toward zero widths.
This complicates fitting widths distributions mainly because small width resonances are more likely to be lost in the noise of an experiment.

For elastic scattering, $\nu_{\mathrm{el}}=1$ and, owing to the strong energy dependence of the neutron penetrability factor, one typically uses reduced width amplitudes to avoid bias.
For capture, in which the compound nucleus can couple to a very large number of states below it, $\nu_{\gamma}$ is assumed to be very large  ($\nu_{\gamma}\rightarrow\infty$). In practice, we may also determine $\nu_{\gamma}$ from a fit to the width distribution, provided detailed capture width data is available.
For fission, $\nu_f$ is observed empirically to lie around 2-3~\cite{Atlas}.

% ------------------------------------------------------------
\subsection{Energy dependence of average resonance parameters}
% ------------------------------------------------------------

The correlations we seek to exploit from RMT rely on knowing the average widths or mean spacings for resonances within a spingroup.
Here we quickly review relevant results.
We will remind the reader that the mean spacing and the average widths vary slowly on the energy scales of the typical resonance width or inter-resonance spacing.
Thus, we can use an entire resonance sequence to determine these parameters without worrying about an energy dependent bias.

\paragraph{Average level spacing - }
Phenomenologically, we know that for light nuclei, the average spacing $\overline{D}$ is of the order of $\sim$ MeV, so there are very few resonances, and our classification algorithm should not be applicable.
A direct fit with R matrix code is the best option and, as there are very few resonances, there is no real need for automation.
For medium mass nuclei, $\overline{D} \sim$ keV, so there are enough resonances to enable robust classification by $L$ and a potential for classification by $J$.
Here we can begin to address poor classification of resonances at high energy that impact neutron capture and leakage.
For heavy nuclei, $\overline{D} \sim$ eV, so there are many resonances very close together.
This is the ideal situation for our classification code.
The average level spacing is inversely proportional to the level density for the corresponding spins and parities.
From consideration of back-shifted Fermi gas models of level density, we expect the  energy dependence of $\overline{D}(E)$ to be rather weak and only noticeable on energy scales of $\sim$ MeV~\cite{Atlas,RIPL3}.

\paragraph{Average neutron width - }
The neutron (or elastic) width of a given resonance is directly related to the reduced width \cite{Atlas,Froehner:2000}
\begin{equation}
\Gamma_{n c} = 2P_c\gamma^2_{n c} = \Gamma_{n c}(|E_n|)\frac{P_L(E)}{P_L(|E_n|)}.
\label{eq:ave-neutron-width}
\end{equation} 
Here the neutron penetrability factor $P_c$ is related to the imaginary part of the logarithmic derivative of the neutron-target relative wavefunction at the channel radius boundary $a_c$ in the R-matrix approach \cite{Froehner:2000}.  In the case of neutron projectiles, the penetrability only depends on the orbital angular momentum $L$.
Thus we have a handle on the average neutron width through the average reduced neutron width 
$\overline{\gamma^2_{\mathrm{el}}}$.

The average reduced neutron width $\overline{\gamma^2_{\mathrm{el}}}$ is independent of the incident energy and all energy dependence of the average neutron width comes from the penetrability factor whose energy dependence is weak on the energy scales of the inter-resonance spacing.
Also, the average reduced width is proportional to the pole strength, 
$s_c=\overline{\gamma^2_{c}}/\overline{D}$,
and, therefore, the neutron strength function, $S_\ell = 2 k_c a_c s_c\sqrt{(1 eV)/E}=2 k_c a_c \frac{\overline{\gamma^2_{c}}}{\overline{D}} \sqrt{\frac{(1 eV)}{E}}$ \cite{Atlas,Froehner:2000}.
Here $k_c$ is the neutron wavenumber and $a_c$ is the channel radius in the R-matrix formalism.
This suggests that we can compute the average width directly from either systematics or using an optical model calculation.
Either way, it varies slowly on the energy scale of interest, so we may take it as constant.
While reduced neutron widths may be slowly varying with energy in accordance with the neutron strength function, the average neutron width decreases with energy on average because of the additional factor of the neutron penetrability.

\paragraph{Average capture width - }
\label{paragraph:capwidth}
The gamma width of a given resonance can be written in terms of a penetrability in a way analogous to neutrons, but using a very different language:
\begin{equation}
    \Gamma_{\gamma XL n}(\textrm{single }\gamma) = \epsilon_{\gamma n}^{2L+1}\gamma_{\gamma XL n}^2.
    \label{eq:single-capture-width}
\end{equation}
Here $\epsilon_{\gamma}$ is the energy of a specific gamma and equals the difference in energy of the resonance $n$ (including the separation energy) and a given state in the residual nucleus and $\gamma_{\gamma XL n}^2$ is reduced width amplitude squared for the particular gamma with multipolarity $XL$ from resonance $n$.  

Unfortunately, it is rare that transitions from a resonance to a given state in the residual are measured.  More often we only measure the total radiative width of a resonance
\begin{equation}
    \Gamma_{\gamma XL n} = \sum_{\gamma} \epsilon_{\gamma n}^{2L+1}\gamma_{\gamma XL n}^2.
    \label{eq:total-capture-width}
\end{equation}
Here the sum runs over all gamma transitions starting from resonance $n$ and having the same multipolarity $XL$.
Thus, while $\Gamma_{\gamma XL n}(\textrm{single }\gamma)$ in Eq. \eqref{eq:single-capture-width} would be distributed by $\chi^2$ distribution with $\nu_\gamma=1$, the same cannot be said for the total radiative widths $\Gamma_{\gamma XL n}$.  Usually the direct average of the measured widths is all that can be determined empirically and the fluctuations in the capture widths are strongly damped.  In these cases, the large number of open capture channels causes $\nu_\gamma\rightarrow\infty$ and the capture width distribution to approaches a delta function.  On the other hand, for closed shell or light nuclei, one may expect $\nu_\gamma$ to be rather small.

Nevertheless, starting from Eq. \eqref{eq:single-capture-width}, one can relate the average gamma width to the gamma strength function in analogy with the neutron strength function~\cite{Atlas}:
\begin{equation}
    \overline{\Gamma}_{\gamma XL} = \overline{D}\epsilon_{\gamma}^{2L+1}f_{XL}.
    \label{eq:ave-capture-width}
\end{equation}
Here $\epsilon_{\gamma}$ is the gamma energy, $XL$ is the gamma multipolarity and $f_{XL}$ is the gamma strength function.
$\epsilon_{\gamma}$ and $f_{XL}$ vary slowly on the energy scale of $\overline{D}$ 
~\cite{Atlas}.

\paragraph{Average fission width - }
The average fission width is expected to be related to the fission barrier penetration probability and in the Hill-Wheeler approach, is estimated to be~\cite{Atlas}
\begin{equation}
    \overline{\Gamma}_f=\frac{\overline{D}}{2\pi}\sum_f\frac{1}{1+\exp(2\pi(V_f-E)/\hbar\omega)}.
    \label{eq:ave-fission-width}
\end{equation}
Here $V_f$ is the fission barrier height and $\omega$ is related to the curvature of the barrier.
For actinides, $\omega$ is typically $\sim 0.5$ MeV and $V_f \sim 5-6$ MeV~\cite{RIPL3} so the average fission width is also slowly varying.
As our understanding of the fission channel is still very much phenomenological, we cannot write the widths in terms of a ``fission penetrability'' factor.

%===================================================================================================================
\section{Recasting spingroup assignment as a machine learning problem}
\label{sec:recast}

We assume we have a collection of $N$ resonances, each one of index $n$ with an associated energy $E_n$, a prior spingroup assignment $(L^\textrm{prior}_n, S^\textrm{prior}_n, J^\textrm{prior}_n)$, and widths associated with each open channel $\Gamma_{el,n}, \Gamma_{\gamma, n}$, and possibly $\Gamma_{f,n}$.  In the language of machine learning, we seek to reclassify the resonances according to labels (in our case, the $L$ or both $L$ and $J$ of a sequence) using a series of quantities which are  built from this collection of resonances which we believe are important in distinguishing characteristics of the data.  These distinguishing characteristics are called features.  In subsection~\ref{subsec:labels}, we describe our use of labels, and in subsection~\ref{subsec:features}, our feature choices.

All classifiers require a training step in order to properly be able to make predictions.
This step can be as simple as fitting a function or a more complex statistical study of the input features.
While the nature of this training is algorithm-dependent, we require test data that can be used to perform this training.
Once the classifier is trained, we use a second set of data to validate the quality of the now-trained classifier.
Subsection~\ref{subsec:training} describes our training data and our training strategy in this initial study.

Each classification algorithm has its own strategy and pros and cons.
In subsection~\ref{subsec:classifier}, we discuss our classifier choice and how we optimize its operation.
%However, our main focus in this initial study is the use of the decision tree algorithm described in subsection~\ref{subsec:decisiontree}.

% -----------------
\subsection{Labels}
% -----------------
\label{subsec:labels}

We seek to assign the quantum numbers $L$ and $J$ (and by extension $S$) to a sequence of resonances.  Collectively we refer to the full set of quantum numbers as the ``spingroup'' of the resonances.
In general, it is much easier experimentally to assign $L$ than $J$.
Often the correct $L$ can be assigned on the basis of the shape of a resonance; this is particularly true of s-wave resonances.
The $J$ quantum number is usually assigned using a shape analysis of the outgoing neutron angular distributions in a scattering experiment, a detailed study of the post capture gamma cascade in a capture experiment, or some other complex and expensive experiment or series of experiments.
To complicate the situation, multiple $J$ are possible for a given $L$, each with no obvious distinguishing characteristic other than the interference pattern between resonances with the same quantum numbers.
As a result, in some situations, we may not have enough information to reliably classify resonances by the $J$ quantum number.

Given this situation and the fact that we are using the classifiers in the \scikitlearn\ package~\cite{scikit-learn}, we either label by $L$ or by spingroup.
We will note below that certain features only make sense when classifying by spingroup as the features require ``pure'' sequences corresponding to resonances with common quantum numbers.
We have considered a multistage approach where we first classify by $L$, then by full spingroup, but this would be outside of the scope of present work and it  is thus not discussed in this paper.

% -------------------
\subsection{Features}
% -------------------
\label{subsec:features}
Table~\ref{table:features} lists the features used for classification in our approach. The overall principle is that we define enough relevant features to characterize the resonances within their spingroups, at the same time that we avoid overloading the classifier with redundant or non-discriminative features.
We have experimented with a much larger feature list \cite{NCSPReport}, and detailed studies using the SHAP metric \cite{SHAP.a,SHAP.b} demonstrated that only an handful of targeted features are needed.
The features \texttt{J\_prior} and \texttt{L\_prior} can be thought of as labels that may be ``overridden'' in the classification process and should be viewed as prior estimates based on experimental inference.

Several of the features test whether a given feature is consistent with a known distribution, otherwise known as Out-Of-Distribution (OOD) tests \cite{OOD}.
These tests require knowledge of feature distributions and, therefore, we exploit the predictions of RMT as discussed above: the tendency of resonances to be relatively even-spaced with a spacing distribution given by the Wigner surmise distribution and the tendency of spacings to follow a ``short-long-short-long'' pattern.  The final group of features exploit the known or expected width distributions of the resonances.

We note the features $d(D_{\mathrm{left}})$ and $d(D_{\mathrm{right}})$  (and by extension $p'(D_{\mathrm{left}})$, $p'(D_{\mathrm{right}})$) are poor proxies for $d(\rho)$, but can be used when classifying by $L$ alone.

\begin{widetext}
\begin{center}
\begin{longtable*}{p{1.5cm}cp{2.0cm}p{1.5cm}p{10cm}}
\caption{
    List of features used by our classifier.  The ``Labels'' column denotes whether the particular feature is used when classifying by $L$ 
    alone (``$L$'') or by the full set of spingroup quantum numbers (``sg'').  The ``Indep. Params.'' column lists the resonance independent 
    parameters needed to compute this feature. Similarly, the ``Dep. Params.'' column lists the parameters of a given resonance 
    (or neighboring resonance) needed to compute this feature.
    \label{table:features}
}\\
\toprule\toprule
\textbf{Feature} & \textbf{Labels} & \textbf{Indep. Params.} & \textbf{Dep. Params.} & \textbf{Description} \\
\midrule
\endfirsthead
\multicolumn{5}{c}%
{\tablename\ \thetable\ -- \textit{Continued from previous page}} \\
\toprule\toprule
\textbf{Feature} & \textbf{Labels} & \textbf{Indep. Params.} & \textbf{Dep. Params.} & \textbf{Description} \\
\midrule
\endhead
\hline \multicolumn{5}{r}{\textit{Continued on next page}} \\
\endfoot
\bottomrule\bottomrule
\endlastfoot

\texttt{L\_prior} & $L$, sg & n/a & $L_n$
           & The orbital angular momentum of the $n^{\mathrm{th}}$ resonance.  Assigned prior to classification.
           \\ \midrule

\texttt{J\_prior} & sg & n/a & $J_n$
           & Total angular momentum of the $n^{\mathrm{th}}$ resonance.  Assigned prior to classification.
           \\ \midrule

\texttt{pos/len} & $L$, sg & n/a & $n/N$
                 & The position $n$ of the resonance within the sequence divided by the length of the sequence $N$.
                   Experimentally, resonances of higher energy are more likely to be misplaced or missed, so
                   this feature is a way to predict whether or not a resonance in a given region of the sequence may be problematic.
                   We note for the training data described below, \texttt{pos/len} does not help as the training data is not biased in this way. \\ \midrule

$d(D_{\mathrm{left}})$ & $L$, sg & $\overline{D}$ & $D_n$
              & The quadratic difference (see Eq. \eqref{eq:distance}) between the $n^{\mathrm{th}}$ spacing and 
                the average spacing, $|D_n-\overline{D}|^2/\overline{D}^2$
                where $D_n=E_{n}-E_{n-1}$. Small values signal OOD. \\ \midrule

$d(D_{\mathrm{right}})$ & $L$, sg & $\overline{D}$ & $D_{n+1}$
               & The quadratic difference (see Eq. \eqref{eq:distance}) between the 
                 $(n+1)^{\mathrm{th}}$ spacing and the average spacing, $|D_{n+1}-\overline{D}|^2/\overline{D}^2$
                 where $D_{n+1}=E_{n+1}-E_{n}$.  Small values signal OOD.  \\ \midrule

$p'(D_{\mathrm{left}})$ & sg & $\overline{D}$ & $D_n$
               & For the current energy $E_n$, this is the signed P-value (see Eq. \eqref{eq:signed-p-value}) for the 
                 spacing between the current energy and the next lower energy, $D_n=E_{n}-E_{n-1}$
                 Small values signal OOD.  In particular, positive values signal missing resonances while small 
                 negative values signal the presence of extra resonances. \\ \midrule

$p'(D_{\mathrm{right}})$ & sg & $\overline{D}$ & $D_{n+1}$
               & For the current energy $E_n$, this is the signed P-value (see Eq. \eqref{eq:signed-p-value}) 
               	 for the spacing between the current energy andthe next higher energy, $D_{n+1}=E_{n+1}-E_{n}$
                 Small values signal OOD.  In particular, positive values signal missing resonances while small 
                 negative values signal the presence of extra resonances. \\ \midrule

$d(\rho)$ & sg & $\overline{D}, \overline{\rho}=-0.27$ & $D_n, D_{n+1}$
               & The quadratic difference between (see Eq. \eqref{eq:distance}) the $n^{\mathrm{th}}$ spacing-spacing 
                 correlation and the expected correlation coefficient,
                 $\overline{\rho}=-0.27$: $|D_nD_{n+1}/\overline{D}^2 - \overline{\rho}|^2/\overline{\rho}^2$.  Small values signal OOD.  \\ \midrule

$d(\Gamma_{el})$ & $L$, sg
                 & $\overline{\Gamma}_{\mathrm{el}}$
                 & $\Gamma_{\mathrm{el}, n}$
                 & The quadratic difference (see Eq. \eqref{eq:distance}) between the $n^{\mathrm{th}}$ elastic 
                   width and the average elastic width, $|\Gamma_{\mathrm{el}, n}-\overline{\Gamma}_{\mathrm{el}}|^2/\overline{\Gamma}_{\mathrm{el}}^2$.
                   Small values signal OOD.  In the future, we will explore the use of the p-value to replace this feature.\\ \midrule

$d(\Gamma_{\gamma})$ & $L$, sg
                     & $\overline{\Gamma}_{\gamma}$ & $\Gamma_{\gamma, n}$
                     & The quadratic difference (see Eq. \eqref{eq:distance}) between the $n^{\mathrm{th}}$  
                       capture width and the average capture width,
                       $|\Gamma_{\gamma, n}-\overline{\Gamma}_{\gamma}|^2/\overline{\Gamma}_{\gamma}^2$
                       Small values signal OOD.
                       We use this rather than p-value because this feature is insensitive to uncertainty or bias in $\nu_\gamma$. \\ % \hline

\end{longtable*}
\end{center}
\end{widetext}

\subsubsection{Out-Of-Distribution Tests}
% ---------------------------------------
We need a mechanism to test %whether a given value  is a key metric for testing 
whether a given value of $x$ is consistent with a given distribution or not.   In other words, an ``out-of-distribution'' (OOD) test \cite{OOD}. 

In the subsequent OOD feature definitions, we adopt the following. For a given Probability Density Function (PDF) $P(x)$ defined on the interval $(x_{\mathrm{min}}, x_{\mathrm{max}})$, we define the Cumulative Distribution Function (CDF)
\begin{equation}
	CDF(x) = \int_{x_{\mathrm{min}}}^x dx' P(x')
\end{equation}
and the Survival Function (SF)
\begin{equation}
	SF(x) = \int_x^{x_{\mathrm{max}}} dx' P(x') = 1 - CDF(x)
\end{equation}

For the OOD tests considered, we use one of four classes of metrics:
\begin{enumerate}
\item Value of the PDF  at $x$, $v(x) = P(x)$

\item P-value
    \begin{equation}
    	\label{eq:p-value}
	    p(x) = \left\{
	    \begin{array}{rl}
		    CDF(x) & x<\overline{x} \\
		    SF(x) & x>\overline{x}
	    \end{array}
	    \right. ,
    \end{equation}
    which gives the probability that a more extreme value of $x$ may be drawn.  Here $\overline{x}$ is the mean of the distribution in question.

\item ``Signed'' P-value
    \begin{equation}
    	\label{eq:signed-p-value}
	    p'(x) = \left\{
	    \begin{array}{rl}
		    -CDF(x) & x<\overline{x} \\
		    SF(x) & x>\overline{x}
	    \end{array}
	    \right.
    \end{equation}
    For the spacing distribution, it is useful to distinguish whether a spacing is too small (indicating an extra resonance is found in the current sequence) or too large (indicating a resonance is missing from the current sequence).

\item Distance to mean, normalized by the mean value to remove the overall scale from the metric
    \begin{equation}
    	\label{eq:distance}
	    d(x) = \left|x-\overline{x}\right|^2/\overline{x}^2
    \end{equation}

\end{enumerate}

\begin{figure}
\includegraphics[width=0.45\textwidth,clip=true,trim=72mm 20mm 88mm 25mm]{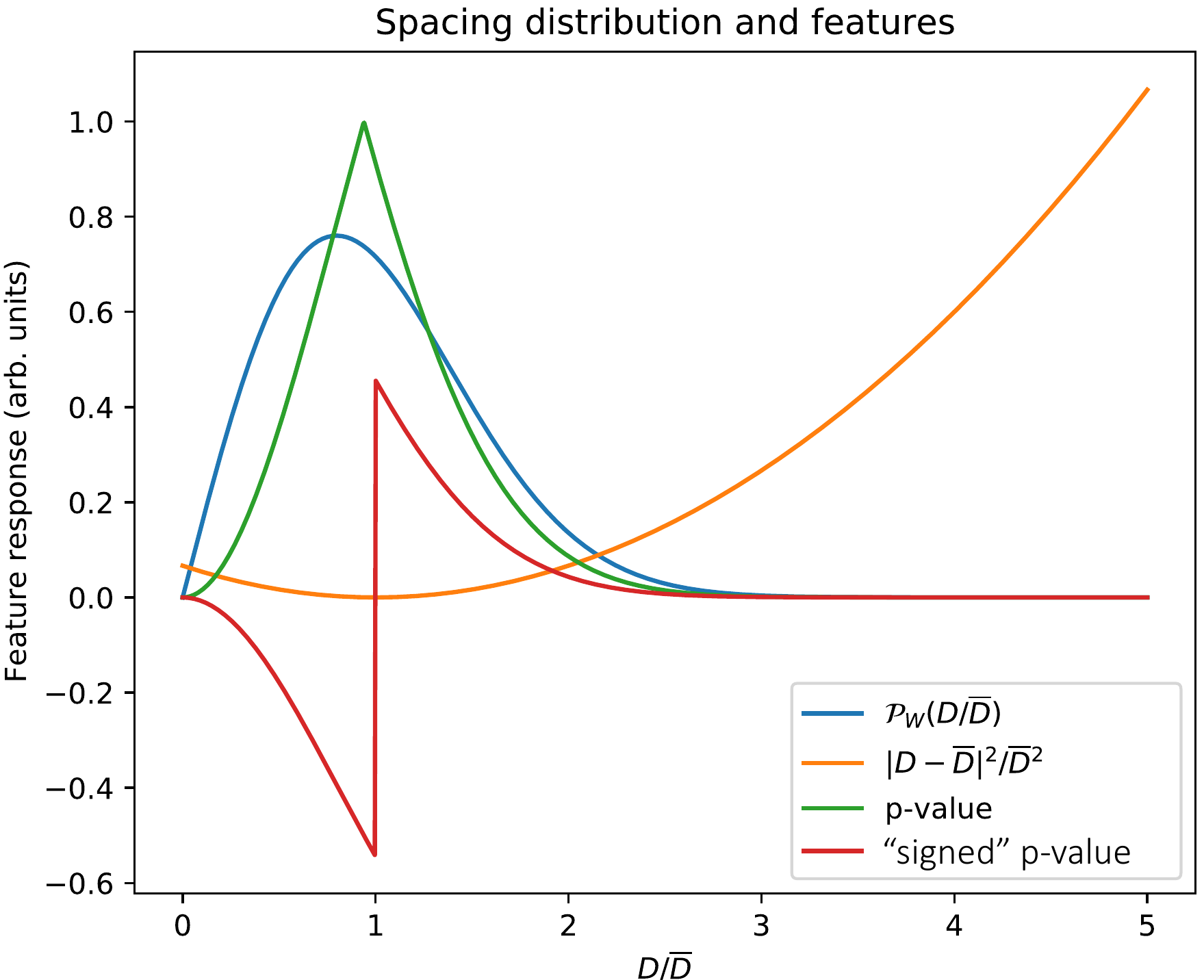}
\caption{
\label{fig:spacing-features}
Spacing distribution OOD features.  The use of the ``signed'' p-value allows us to distinguish between overly small spacings (indicating one or more extra resonances in the sequence) and overly large spacings (indicating one or more missing resonances from the sequence).}
\end{figure}

\begin{figure*}
\begin{center}
\includegraphics[width=0.45\textwidth]{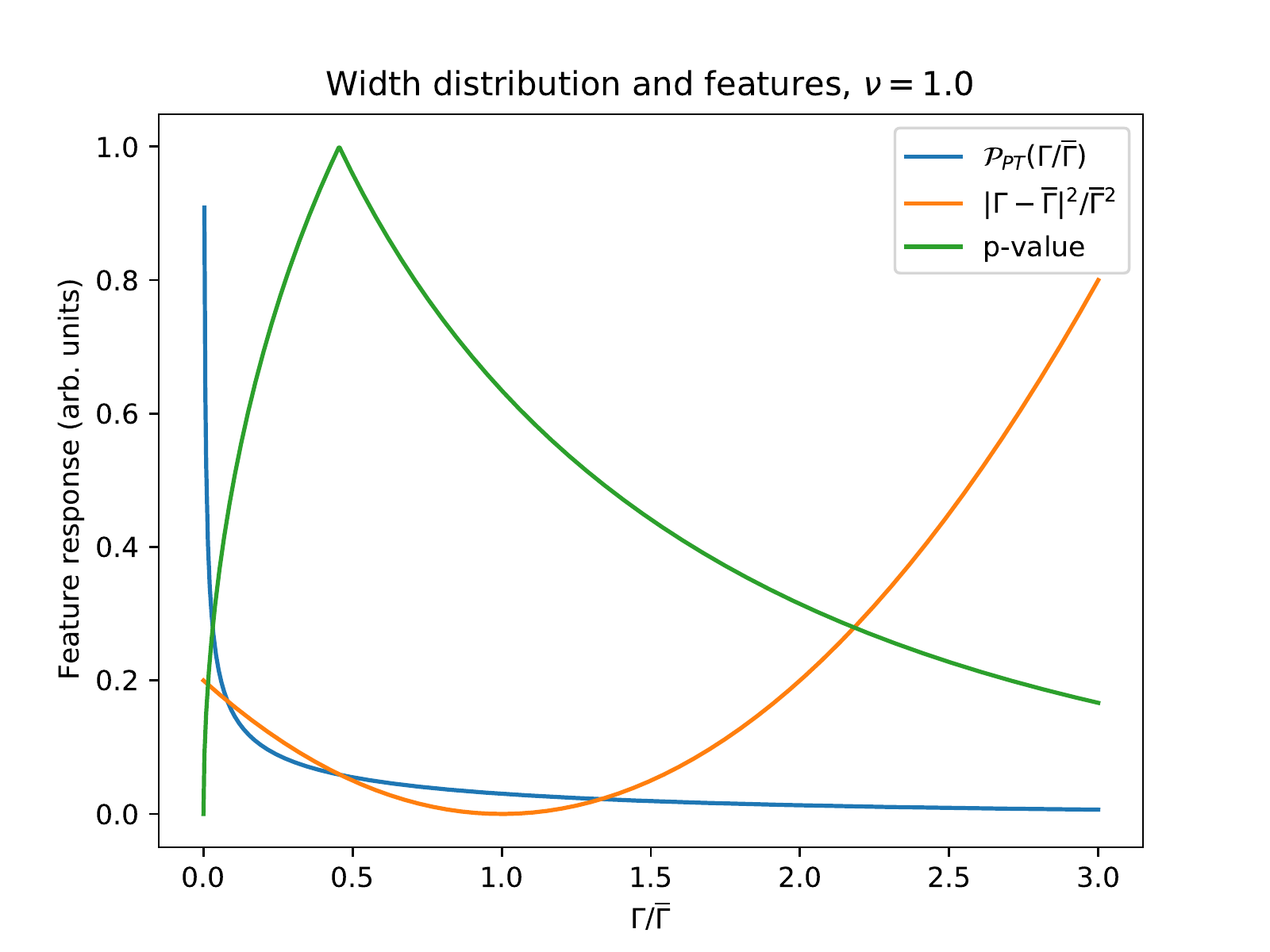}
\includegraphics[width=0.45\textwidth]{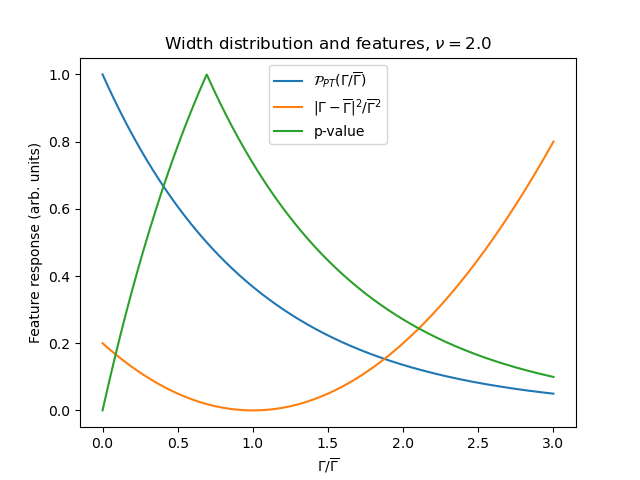}
\includegraphics[width=0.45\textwidth]{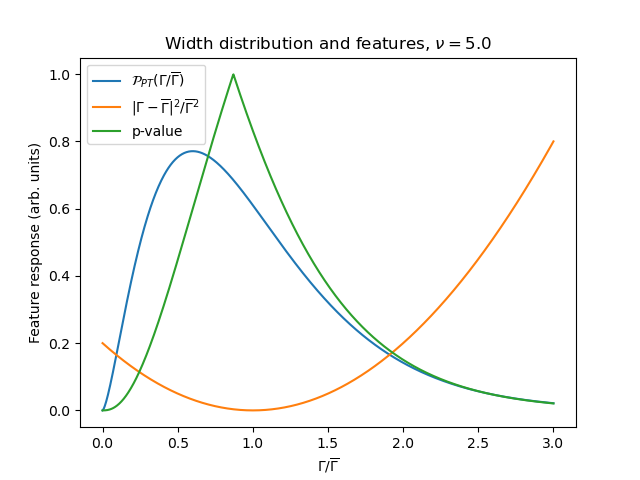}
\includegraphics[width=0.45\textwidth]{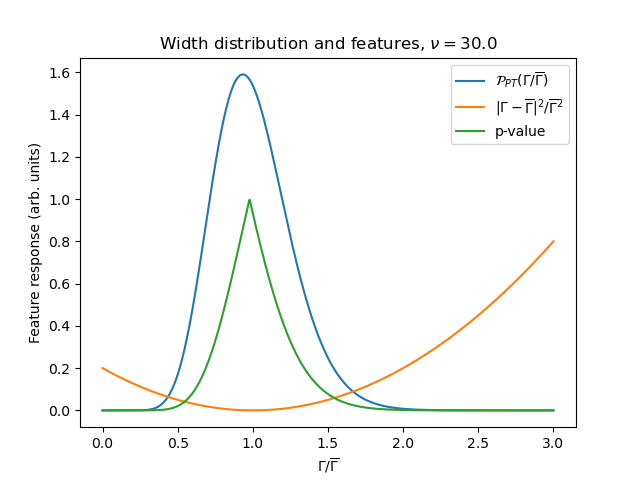}
\end{center}
\caption{
\label{fig:width-features}
Width distribution OOD features.  The Porter-Thomas distribution is divergent at small widths for $\nu=1$ so it is a poor choice for OOD testing.}
\end{figure*}

The various OOD testing features are illustrated in Figures \ref{fig:spacing-features} and \ref{fig:width-features}.  In other Extreme Value Testing (EVT) methods, one assigns a criteria for an OOD data point (say more than 3-sigma).  Here we use the OOD metric as a feature and provide properly labeled training data so that the classifier can learn what criteria should be used for OOD detection.

\subsubsection{Spacing features}
% ------------------------------
Several feature distributions require a predetermined value of the average spacing $\overline{D}_{LJ}$ for each spingroup.
We can achieve this several ways:
\begin{enumerate}
    \item Direct averaging of spacings
    \item Fit the cumulative level distribution to extract $1/\overline{D}$ %\fixme{not sure cumulative level distribution is described/defined}
    \item Fit the nearest neighbor spacing distribution to Wigner's surmise distribution
    \item Take values from a pre-existing compilation
\end{enumerate}
Options \#1-3 can be performed as an initial training step for our classifiers or even iteratively improved as we reclassify resonances.
Both \#2 and \#3 can be achieved by fitting empirical distributions (either cumulative level distribution for \#2 or cumulative Wigner surmise distribution for \#3).
We note that the breadth of the Wigner surmise means that \#3 converges slowly as the number of spacings increases.
%All of these methods can be refined by using the post-classification spin group sequences as input for another round of $D$ extraction.

Options \#1 and \#2 may also be used if one does not have robust $J$ assignments to determine $\overline{D}_L$.
Simple consideration of the number of energies on a given interval leads one to the follow sum rules for the resonance spacings for the full sequence $\overline{D}$, the subsequence of resonances with a given orbital angular momentum $\overline{D}_L$ and the subsequence of resonances within a spingroup $\overline{D}_{LJ}$:
\begin{equation}
	\frac{1}{\overline{D}} = \sum_L \frac{1}{\overline{D}_L}
  \label{eq:Dcombination}
\end{equation}
and
\begin{equation}
	\frac{1}{\overline{D}_L} = \sum_J \frac{1}{\overline{D}_{LJ}}.
  \label{eq:DLcombination}
\end{equation}

\subsubsection{Width features}
% ----------------------------
Many features distributions require knowledge of the average width $\overline{\Gamma}$ and number of degrees of freedom parameter $\nu$ of the appropriate Porter-Thomas distribution.  We will approach each width and $\nu$ pair the same way.
As a technical aside, small-width resonances tend to be missed experimentally, and we need a method for determining these widths that is robust against this bias. When determining the average widths, we fit the width survival function of the Porter-Thomas distribution.  By integrating from large to small widths, the dominant part of the integral comes from the region in widths that are most accurately determined experimentally. This also can be used to yield the $\nu$ for the fission channels and the total width.

For elastic reactions, $\nu$ is assumed to be unity when classifying by spingroup or the number of allowed $J$ values when classifying by $L$.  Also, when fitting elastic width distributions, we can either fit the experimental width distribution or the reduced neutron width distribution.  We note that the presence of doorway states may distort the neutron width distribution \cite{doorway}.  We may explore this effect in future work.

For neutron capture, the width distribution is often very narrow and $\nu_\gamma\rightarrow\infty$.
In this case, it is appropriate to directly average the capture widths.
We note that in many older data sets, the capture widths were assigned based on the average widths which can introduce serious bias in the classification.
To counter this bias, we implemented in our codes the option to ``turn off'' capture widths as an active feature.
When the distribution is not so narrow, we must approach the capture distribution in the same manner as elastic or fission widths.

% -------------------
\subsection{Training}
% -------------------
\label{subsec:training}

Supervised machine learning algorithms, such as those used in this work, rely on having a large amount of labeled data for training purposes.
With this training data, the machine learning algorithm will ``learn'' the solution physics, without a need for an explicit solution formulation.
While experimental resonance data might be used for training, there are several problems with such approach:
\begin{itemize}
	\item the number of resonances available for a given nucleus are often only on the order of hundreds of resonances, on the borderline of what is needed for robust training
	\item experimental data is not guaranteed to have correct labeling by either $L$ or spingroup
	\item experimental data may be missing smaller resonances or have ``contamination'' by resonances from other nuclei in the target or surrounding experimental apparatus.
\end{itemize}
Compilations,  such as the {\em Atlas of Neutron Resonances}~\cite{Atlas} and/or evaluations such as the ENDF library~\cite{ENDF-VIII.0}, are attractive sources of training data, but even these do not always have enough statistics and/or are not guaranteed to have correct labeling either.
Thus, we are forced to consider synthetic training data.

Synthetic data can be constructed in a way nearly indistinguishable from real data and
can be generated from the well understood statistical properties of nuclear scattering physics described in Section~\ref{subsec:rmt}.
In Ref.~\cite{mcres}, the authors describe the addition of a stochastic resonance generator to the FUDGE processing system~\cite{FUDGE}.
This tool takes advantage of many known results from GOE random matrices~\cite{Weidenmueller:2009,Mitchell:2010,Mehta:1967}:
\begin{itemize}
\item Realizations are GOE consistent by construction since a GOE Hamiltonian matrix is generated as the first step in making a resonance realization and the eigenvalues of this matrix provide the resonance energies.
\item The eigenvalues of this matrix are not quite the resonance energies, since the mean level spacing $\overline{D}$ is incorrect.
      We rescale the eigenspectrum so that the mid-range of the spectrum's level spacing matches the required $\overline{D}$.
\item The widths are drawn from a Porter-Thomas distribution as in traditional ladder generators found in nuclear data processing codes.
\item The reconstructed pointwise cross sections generated from this resonance realization can be generated using any level of approximation to the R-matrix.
      Although we could use the Reich-Moore approximation as it is generally regarded as the most appropriate and accurate approximation for nuclei with $Z > 10$, we do not need the reconstructed cross section for this project.
\end{itemize}

In order to simulate the quantum number misassignments seen in the real world,
we randomly misassigned a fraction of the resonances in these synthetic sequences.
The fraction of reassigned resonances can be varied to test the reliability of our method.
Because such  reassignments occur independent of either resonance energy or width, they do not currently fully mimic actual experimental effects.
We also do not consider other experimental effects, such as resonance energy shifts caused by moderation in the neutron source, Doppler broadening in the target, or target  contamination.  These and other effects impact the initial resonance quantum number assignments in an uneven way - in a shape analysis
$L=0$ resonances are easy to identify but higher $L$ resonances have less certain assignments at higher energies.
Other methods of spingroup assignments have their own biases.
We will explore these experimental impacts in future works.
We have considered adding additional metadata to each resonance to help the classifier understand the quality of the spingroup assignment, and this is another topic for a future work.

In this first incarnation of a machine learning tool, we used the \scikitlearn\ \texttt{test\_train} function~\cite{scikit-learn}  to split input data into training data and test data.
The fraction of data randomly selected for training, with the remaining input data reserved for testing, can be chosen through a parameter in the function call. 
In the future, we aim to improve the training regimen using a combination of expert knowledge and numerical experimentation.

% ---------------------
\subsection{Classifier}
% ---------------------
\label{subsec:classifier}

The approach presented in Section~\ref{subsec:labels} defines labels, and the approach described in Section \ref{subsec:features} converts sequences of neutron resonances into sets of features which can be coupled into any machine learning classifier.
As the main focus of the work is on the methodology of spin classification of neutron resonances through machine learning, we employed pre-packaged ML classifiers from \scikitlearn~\cite{scikit-learn}.
While we performed a preliminary assessment of different classifiers and associated hyperparametrizations, we illustrate the approach with a Multi-Layer Perceptron classifier.   Multi-Layer Perceptrons belong to the family of Neural Network algorithms. 

This assessment with multiple classifiers and hyperparametrizations was done in a preliminary fashion, using only training and test data sets, with bias mitigated through multiple training events. Ideally, however, independent validation sets should be used in a rigorous optimization and/or using approaches such as $K$-fold cross-validation \cite{k-fold}. Nevertheless, the choice of classifier and hyperparameters should not significantly impact the conclusions presented in this work, and the results should be transferable to other choices of classifier and hyperparametrizations. Now that the proof of principle is established in this work, we leave the optimization step for a future work.

\subsubsection{Multi-Layer Perceptron}
\label{sec:MLP}

As with other supervised learning algorithms, the Multi-Layer Perceptron (MLP) ``learns'' a function  that defines a hyperplane that optimizes the separation of data points with different labels. One difference from other ML algorithms, such as logistic regression, for example, is that in MLP, there can be one or more non-linear layers, called hidden layers, between  the input and the output layers~\cite{Rumelhart:1986,scikit-learn}. The learning process is done by training on a dataset, whose data is characterized by a set of features, for which the labels are known. The training uses  backpropagation~\cite{MLP_backpropagation,backpropagation_1,backpropagation_2}, which adjusts the weights in each hidden layer to approximate the non-linear relationship between the input and the output layers.

While the MLP can learn a non-linear function approximator for either classification or regression, we use the \texttt{MLPClassifier} function from \scikitlearn~\cite{scikit-learn}  solely for classification.  Our MLP has the number of  non-linear hidden layers as an input hyperparameter and optimizes the log-loss function using the L-BFGS solver for weight optimization~\cite{LBFGS}. The L-BFGS is an optimizer based on  quasi-Newton methods which  approximates the Broyden-Fletcher-Goldfarb-Shanno algorithm (BFGS), requiring significantly less memory. For  smaller datasets, L-BFGS is expected to converge faster and with a better performance~\cite{scikit-learn} than alternatives, such as Stochastic Gradient Descent (SGD)~\cite{sgd} or Adam~\cite{adam}. Our MLP trains iteratively since at each step the partial derivatives of the loss function with respect to the model parameters are computed to update the parameters, with the maximum number of iterations also being a model hyperparameter. In our calculations, we ensured convergence relative to the maximum number of iterations. The strength of the L2 regularization term, which is divided by the sample size when added to the loss, can be used to avoid overfitting by introducing a penalty term in the loss function. Apart from those aforementioned hyperparameters, we assumed \scikitlearn\ default values for all other parameters. Performance could likely be improved by testing different classifiers and by performing a grid search to optimize the hyperparametrizations. As a matter of fact, preliminary investigations in that direction have been done by the authors. However, the scope of this current work is to define and present the method as a proof-of-principle. We, therefore, leave to present optimization efforts for a future publication.

The classifiers from  \scikitlearn\ are set up to randomly split the input data into training and testing subsets. The algorithm is trained only on the training set while the testing one serves as a somewhat independent test of the quality of the training process. Because the splitting of data points (resonances) is random, the classifier is trained in each run with a different training data set, leading to slightly different predictions.  We define a \emph{training seed} as the particular training set obtained through a given random split, and a \emph{training event} as each pass of the input data through the training process, which includes the random split into a training seed and complementary testing subset, defining slightly different classifiers. For this reason, in the application of the method shown in Section~\ref{sec:application}, we define an averaged classifier by averaging the performance and predictions of many different training events, each with different training seeds.

%===================================================================================================================
\section{Application to $^{52}\mathrm{Cr}$}
\label{sec:application}

To assess the efficacy of our approach, we applied our method to the analysis of the $^{52}$Cr resonances from the most recent evaluation for chromium isotopes~\cite{Nobre_2021}.
The average resonance parameters are presented in Table \ref{table:52Crparams_real}.  
$^{52}$Cr has ground state $0^+$ spin/parity so it has five spingroups for $0\leqslant L\leqslant 2$.  

The $^{52}$Cr resonance evaluation in Ref. \cite{Nobre_2021} is taken from the ENDF/B-VIII.0 evaluation published in Ref. \cite{ENDF-VIII.0} and described in 
Leal \etal \cite{ND2010_Cr}.  
The Leal \etal evaluation is a Reich-Moore fit using SAMMY \cite{SAMMY} to a combination of published and unpublished data from ORELA.
Below 100 keV, the fit relied on $^{\mathrm{nat}}$Cr (83.789\% $^{52}$Cr) data of Guber \etal \cite{Guber_2011}.  
Above 100 keV, the evaluation relied on unpublished high resolution transmission data of Harvey \etal on a pair of enriched $^{52}$Cr samples.
No neutron capture or scattering angular distribution data were available above 600 keV, therefore, above 600 keV, the spingroup assignments in Ref.~\cite{ND2010_Cr} are purely based on a shape analysis and evaluator judgement.  
Neither data set used in Ref. \cite{ND2010_Cr} were used in the \textit{Atlas of Neutron Resonance} compilation \cite{Atlas}.

The ENDF/B-VIII.0 evaluation extends from 10$^{-5}$ to 1.450 MeV.  
Above 1.450 MeV, resonances were included mainly to provide background and interference effects to the resonances below 1.450 MeV.  
This is a common practice in ENDF evaluations and is done to ensure an accurate representation of the reconstructed cross section over the given energy region.  

\begin{table*}
  \caption{\label{table:52Crparams_real}
  Average resonance parameters extracted from the actual ENDF/B-VIII.0 resonance evaluation \cite{Nobre_2021}.
  Fitting the cumulative level distribution for all resonances with a given L, we find $\overline{D}_0=31.686 \pm 0.280 $ keV, $\overline{D}_1=11.748 \pm 0.089$ keV and $\overline{D}_2=8.003 \pm 0.028 $ keV, consistent with the above values.}
  \begin{tabular}{cccccccc} \toprule\toprule
    L & J   & Num. res.
            & $\overline{D}$ (keV)
            & $\overline{\Gamma}_n$ (eV)
            & $\nu_n$
            & $\overline{\Gamma}_\gamma$ (eV)
            & $\nu_\gamma$\\
    \midrule
    0 & 1/2 & 69 & 31.69 $\pm$ 0.28 &  13839. $\pm$ 210. &  1.131 $\pm$ 0.020 &  1.2036 $\pm$ 0.0030 &  411 $\pm$ 23 \\
    1 & 1/2 & 46 & 28.72 $\pm$ 0.80  &  485. $\pm$ 41.  &  0.769 $\pm$ 0.066  & 0.4332 $\pm$ 0.0041  &  211 $\pm$ 99\\
    1 & 3/2 & 75 & 19.47 $\pm$ 0.12  &  308.9 $\pm$ 6.8  &  0.994 $\pm$ 0.024  & 0.44647 $\pm$ 0.00046  &  $(16.9 \pm 4.4)*10^3$\\
    2 & 3/2 & 58 & 22.38 $\pm$ 0.24  &  376. $\pm$ 13.  &  0.676 $\pm$ 0.016  & 0.5331 $\pm$ 0.0082  &  40.3 $\pm$ 6.7 \\
    2 & 5/2 & 126 & 11.23 $\pm$ 0.10  &  259.3 $\pm$ 4.5  &  1.098 $\pm$ 0.021  & 0.6381 $\pm$ 0.0022  &  $(5.4\pm 2.0)*10^3$\\
\bottomrule\bottomrule
  \end{tabular}
\end{table*}

\begin{figure*}[!htbp]
 \centering
\includegraphics[scale=0.48,keepaspectratio=true,clip=true,trim=0mm 0mm 0mm 0mm]{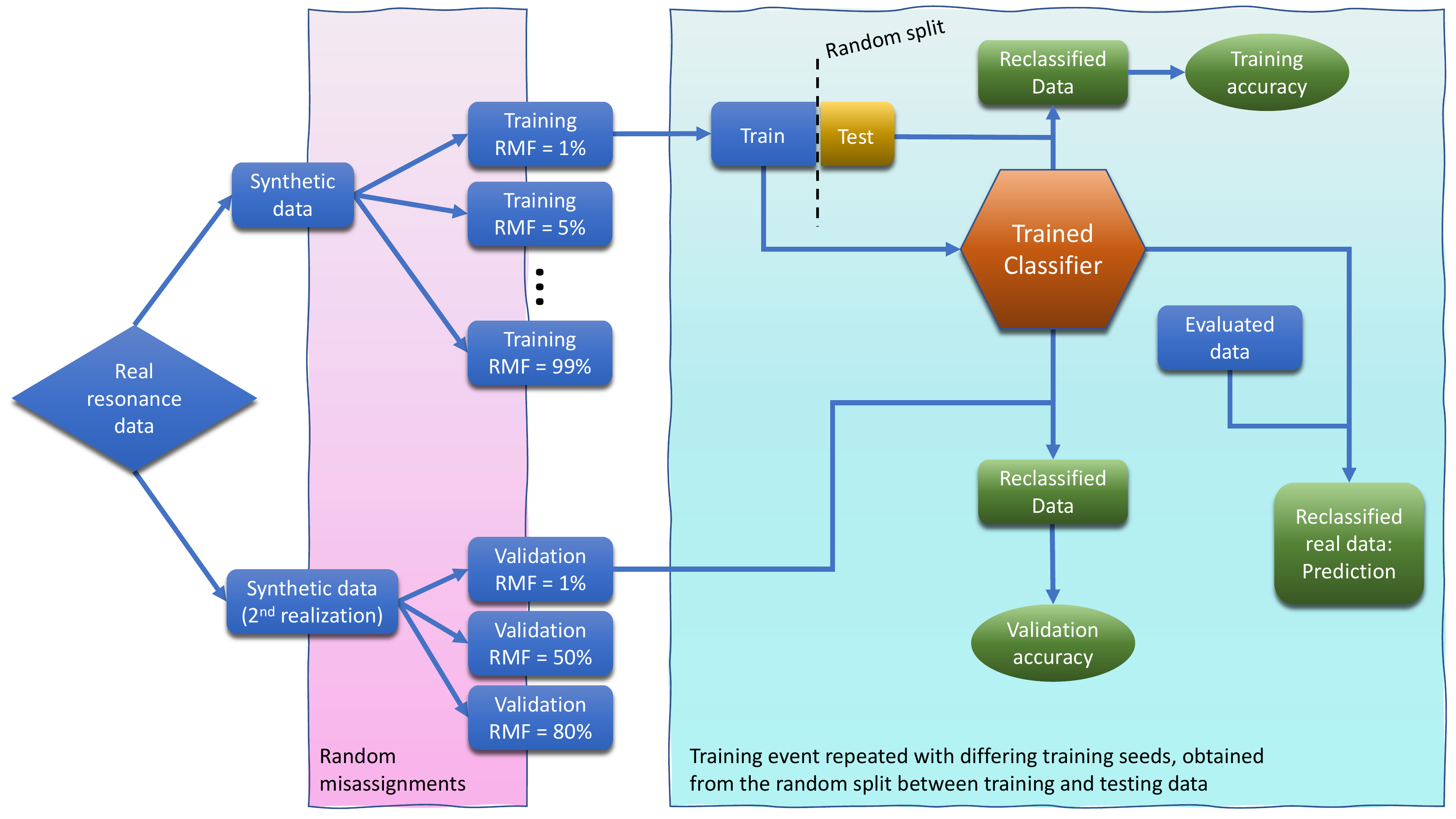}
 \caption{Flowchart illustrating the Machine-Learning approach adopted in this work to re-classify \nuc{52}{Cr} neutron resonances.}
\label{fig:flowchart}
\end{figure*}

In order to illustrate the approach adopted in the current work, which will be described in detail in the following sections, and to facilitate its understanding, we present in Fig.~\ref{fig:flowchart} a flow chart summarizing the steps taken. The reader is encouraged to use Fig.~\ref{fig:flowchart} as a guiding reference while reading the text that will follow.

\subsection{Training with synthetic data}
\label{sec:train_synthetic}

We generated a train/test set in accordance with the methods in Section \ref{subsec:training}.  
The train/test simulated data consists of 4,823 resonances over an energy range $0-20$ MeV.  
In Table \ref{table:52Crparams} we list the spingroups taken from the ENDF/B-VIII.0 evaluation and the average parameters in the train/test set that correspond to the ENDF/B-VIII.0 spingroups.  
We note that although the $\nu_\gamma$ is known for each ENDF/B-VIII.0 spingroup, we assume that $\nu_{\gamma}\rightarrow\infty$ in our train/test data set.

\begin{table}
  \caption{\label{table:52Crparams} The parameters used in our test/train simulated resonance set.
  Average resonance parameters are similar to those extracted from the ENDF/B-VIII.0 resonance evaluation \cite{Nobre_2021}.
  The average spacings below give $\overline{D}_0=34.9$ keV, $\overline{D}_1=12.0$ keV and $\overline{D}_2=7.72$ keV, with $\overline{D}$=4.14 keV.}
  \begin{tabular}{cccccccc} \toprule\toprule
    L & J   & Num. res.
            & $\overline{D}$ (keV)
            & $\overline{\Gamma}_n$ (eV)
            & $\nu_n$
            & $\overline{\Gamma}_\gamma$ (eV)
            & $\nu_\gamma$\\
    \midrule
    0 & 1/2 &  583 & 34.9 & 10$^4$ & 1 & 2.0   & $\infty$ \\
    1 & 1/2 &  673 & 30   & 406.0 & 1 & 0.58  & $\infty$ \\
    1 & 3/2 &  969 & 20   & 303.0 & 1 & 0.56  & $\infty$ \\
    2 & 3/2 &  826 & 24   & 299.0 & 1 & 0.63  & $\infty$ \\
    2 & 5/2 & 1772 & 11.4 & 329.0 & 1 & 0.69  & $\infty$ \\
    \bottomrule\bottomrule
  \end{tabular}
\end{table}

\begin{figure*}[!htbp]
 \centering
\includegraphics[scale=0.37,keepaspectratio=true,clip=true,trim=0mm 2mm 4mm 0mm]{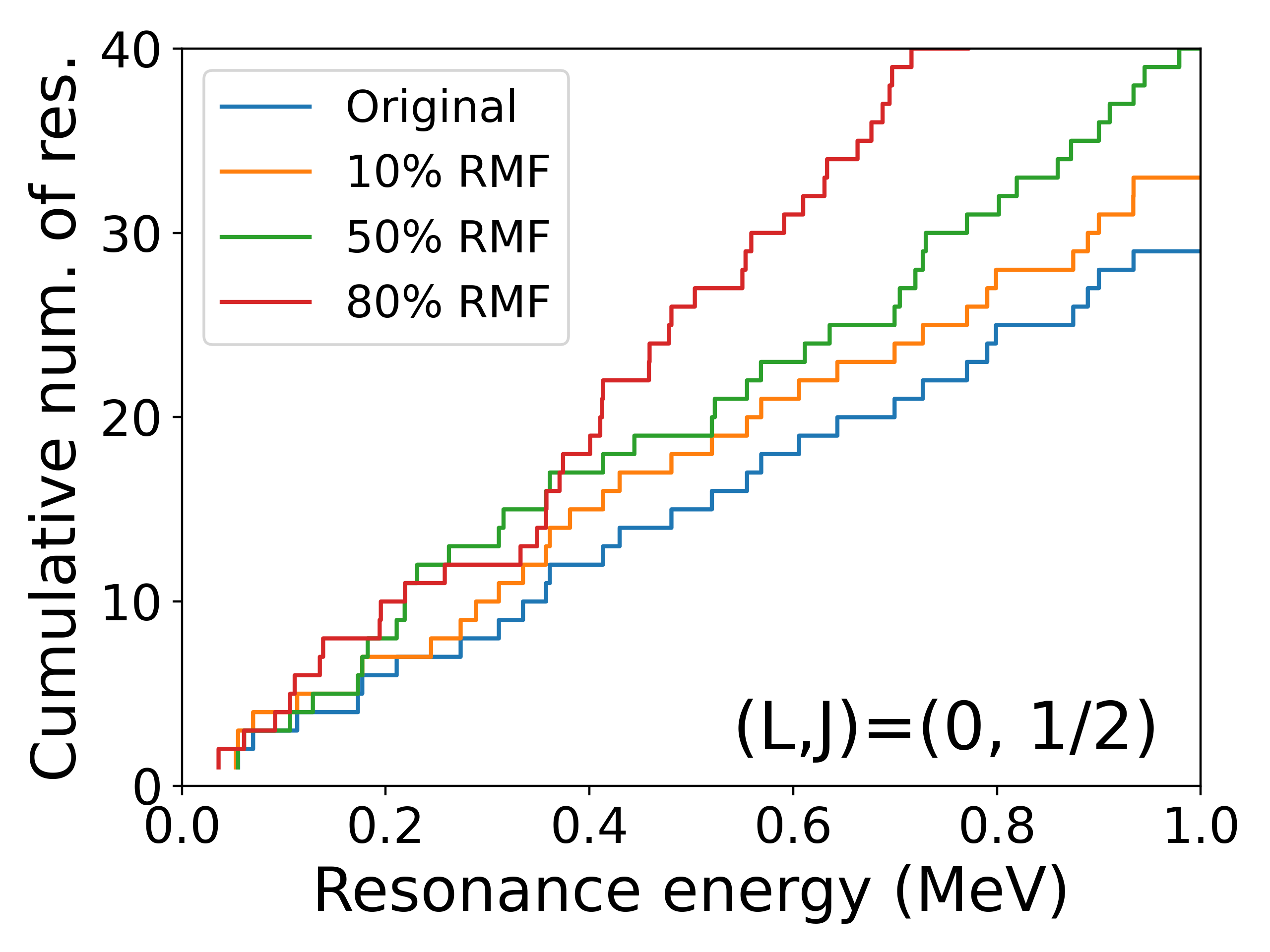}
\includegraphics[scale=0.37,keepaspectratio=true,clip=true,trim=0mm 2mm 4mm 0mm]{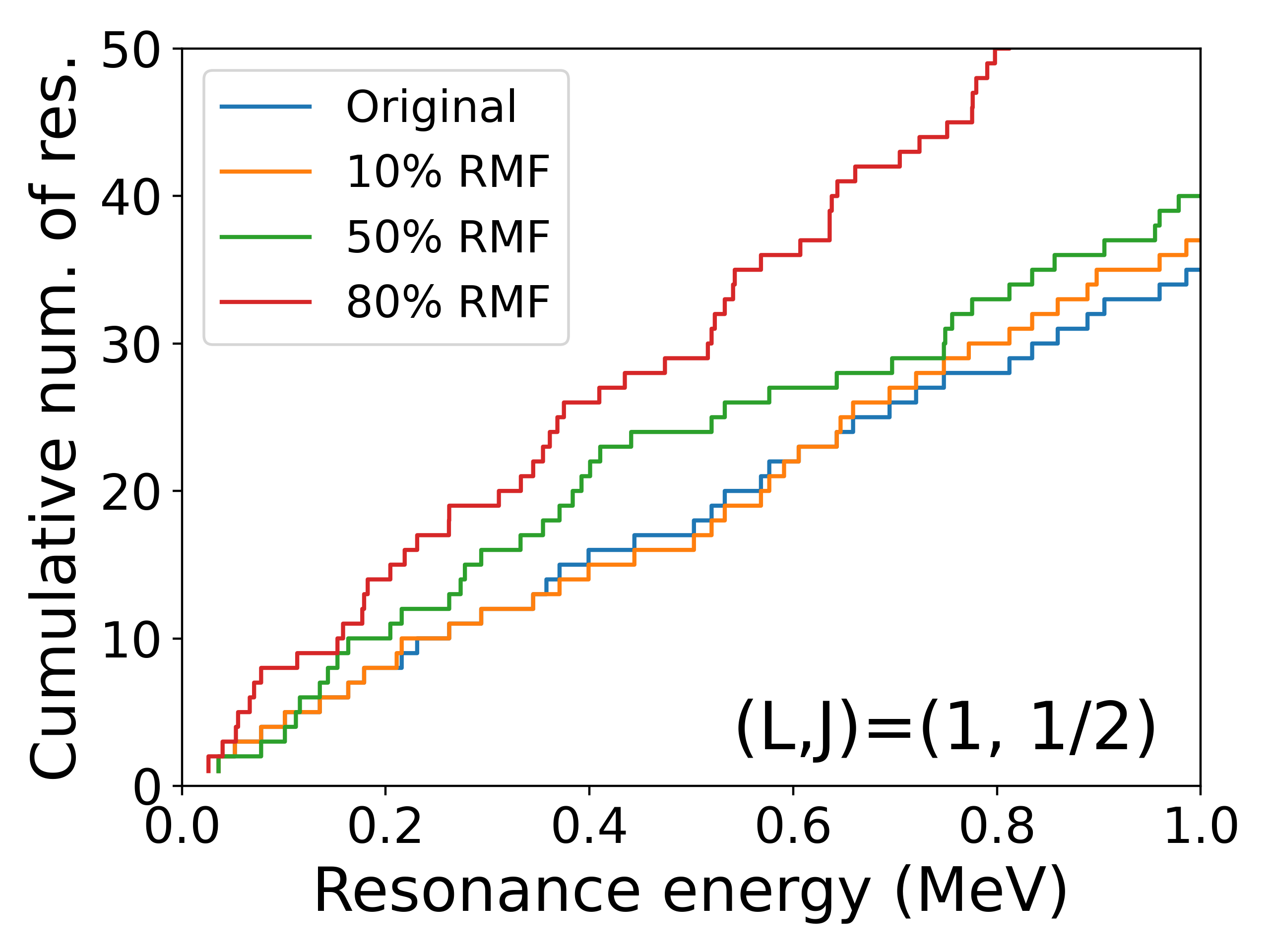}
\includegraphics[scale=0.37,keepaspectratio=true,clip=true,trim=0mm 2mm 4mm 0mm]{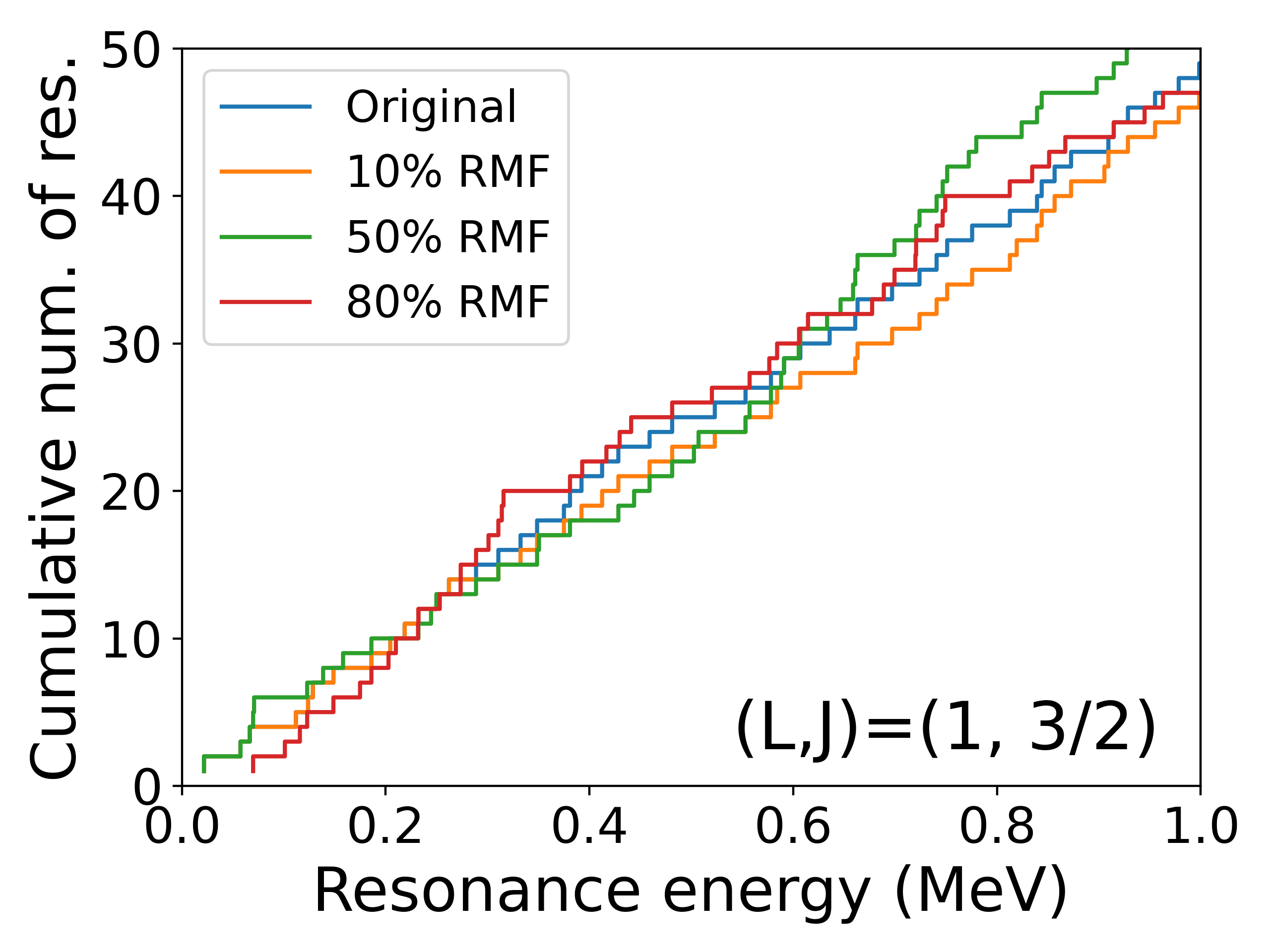}
\\
\includegraphics[scale=0.37,keepaspectratio=true,clip=true,trim=0mm 2mm 4mm 0mm]{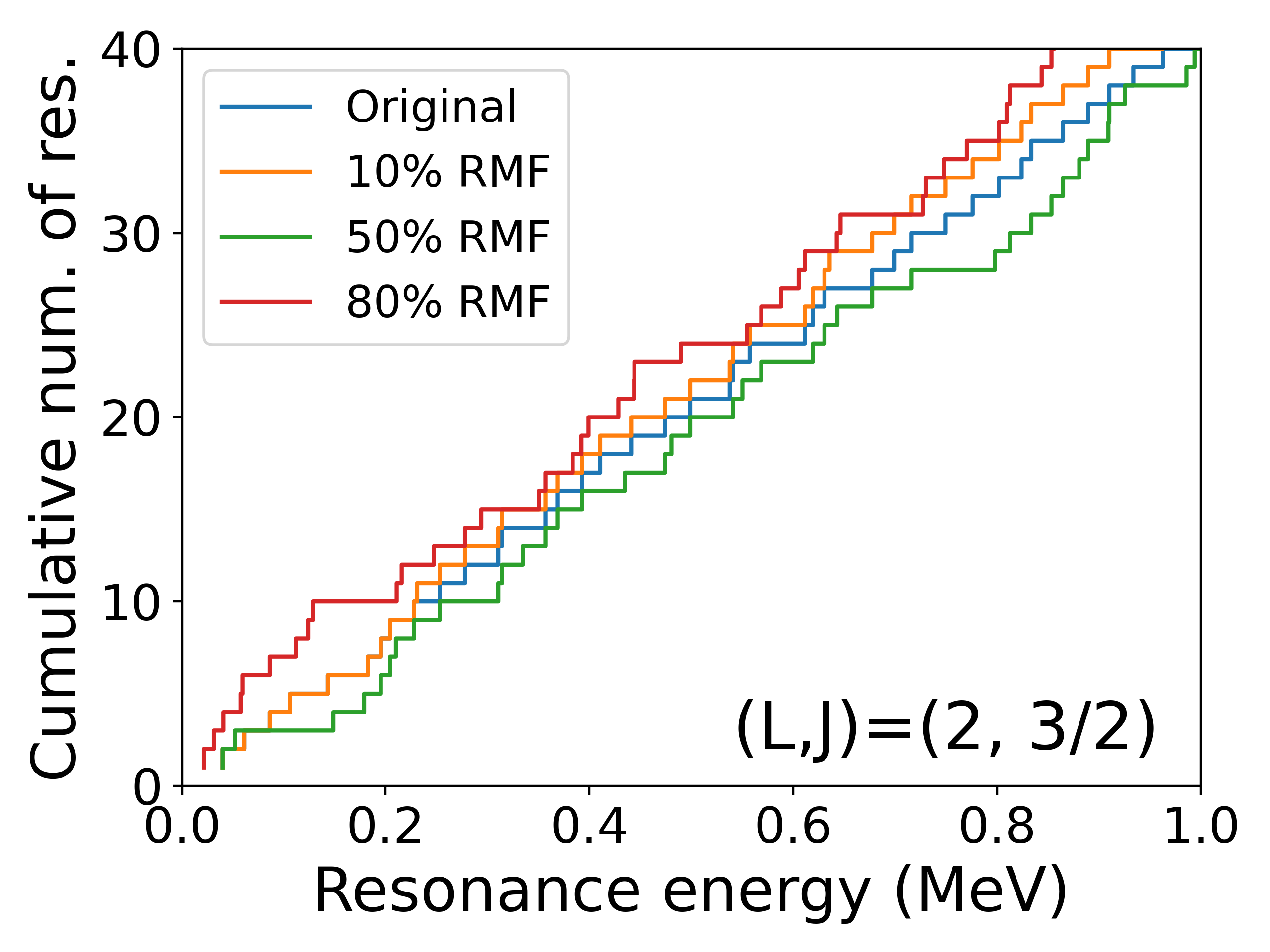}
\includegraphics[scale=0.37,keepaspectratio=true,clip=true,trim=0mm 2mm 4mm 0mm]{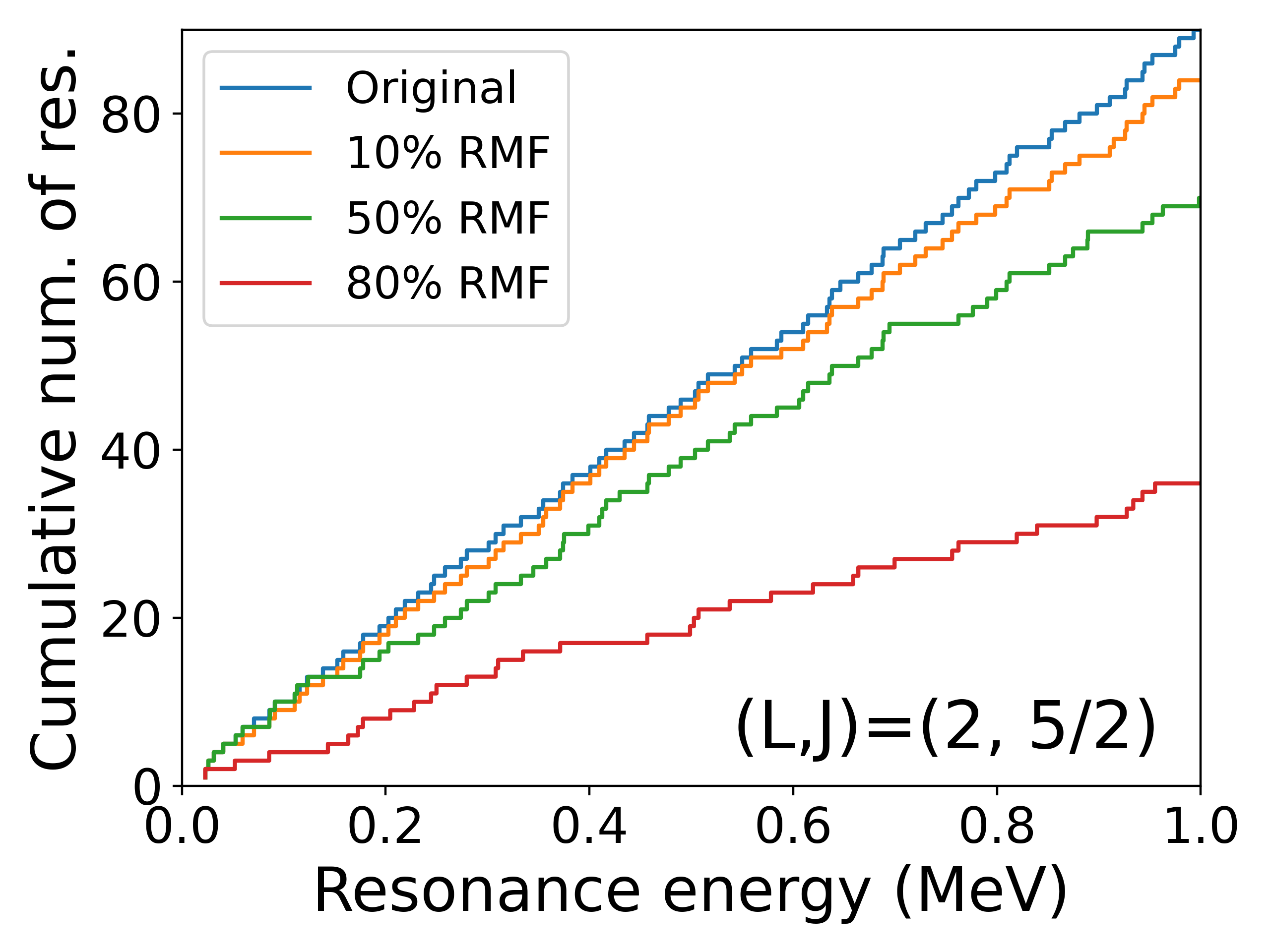}
 \caption{Cumulative number of resonances for synthetic resonance sequences based on \nuc{52}{Cr} for all the spingroups allowed up to $L_{\mathrm{max}}$.}
\label{fig:cr52-cld}
\end{figure*}

\begin{figure}[!htbp]
 \centering
\includegraphics[scale=0.48,keepaspectratio=true,clip=true,trim=0mm 2mm 4mm 0mm]{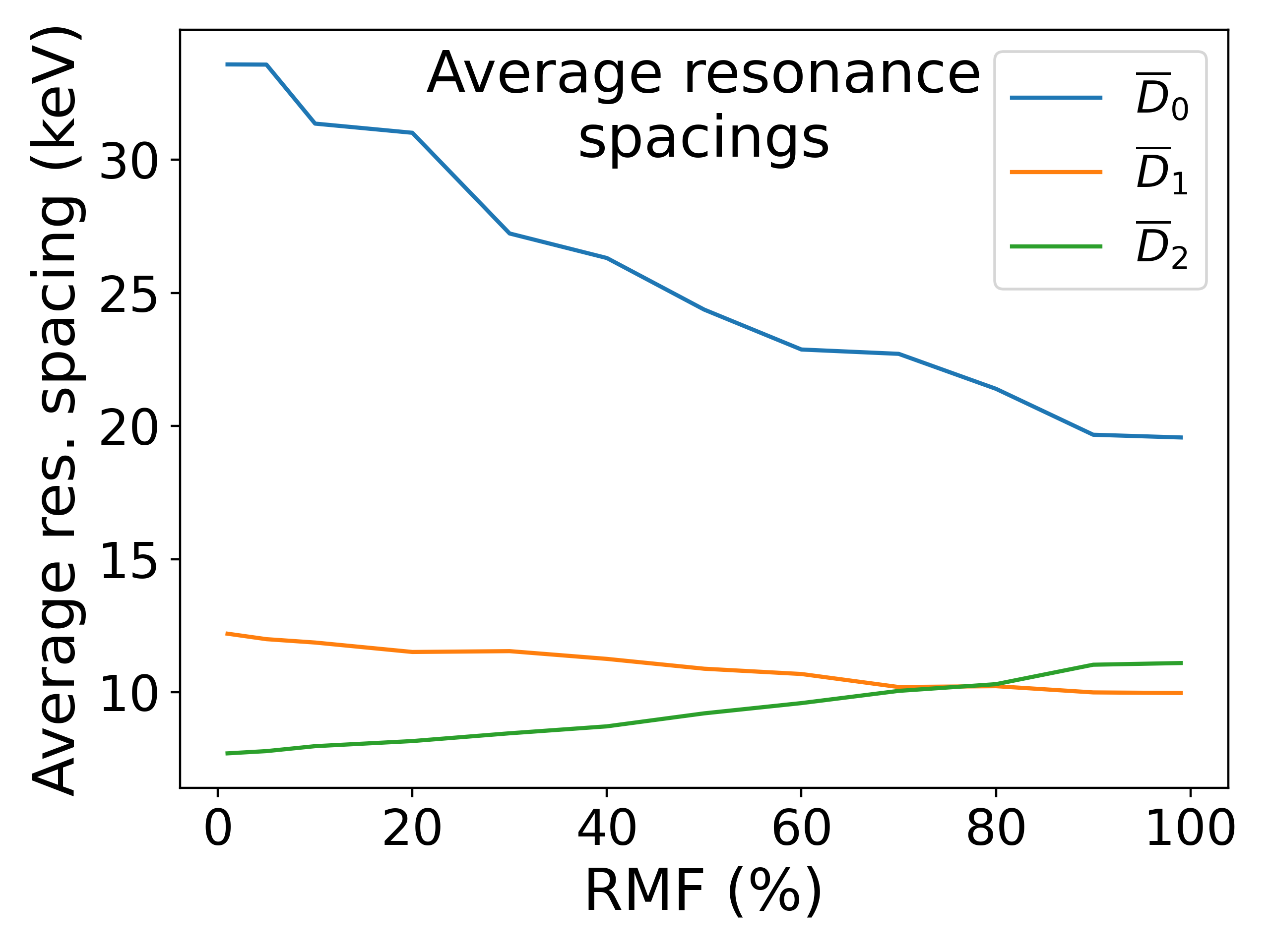}
 \caption{Average spacings $\overline{D}_L$ for different values of $L$ ($L$=0,1, and 2). As shown, the average spacings can vary significantly as a function of the sequence RMF.}
\label{fig:cr52-spacings}
\end{figure}

To simulate the misassignments seen in real data, we randomly misassign resonances in the train/test set in accordance to the prescription in subsection \ref{subsec:training}.  
In Fig. \ref{fig:cr52-cld}, we show the cumulative level distributions for the $L\leqslant2$ spingroups for the original simulated set and three different levels of random misassignment.  
In the following, we refer to the fraction of resonances that receive a random misassignment as the Random Misassignment Fraction (RMF). In each case, we extract the average spacing for the simulated sets.  
In Fig. \ref{fig:cr52-spacings}, we show the extracted average spacing for each $L$ after combining the spacings from each spingroup in accordance with Eq. \eqref{eq:DLcombination}. 
We note that as the degree of misassignment increases, each spingroup's average spacing tends to the global average value of $\overline{D}_{\mathrm{sg}} = 5*4.14$ keV $= 20.7$ keV.  
Thus, the extracted average spacing tends to $20.7$ keV for $L$=0 and $10.4$ keV for both $L$=1 and 2.

With these sets, we trained a MLP algorithm, 
employing the L-BFGS solver, regularizer $\alpha$ set to 1.0 and  maximum number of iterations set to 2000 with 20 hidden layers. Unless noted otherwise, the results shown consider 50 training events. Each training event corresponds to the training of the classifier using one random training seed, using the complementary testing dataset for benchmarking the training. In each synthetic set used for training, we randomly reserve 60\% of the data points in each training event for the actual training while 40\% is used for testing, as explained in Sec.~\ref{sec:MLP}. This is done as a way to assess the quality of the training process, or how well the algorithm can be trained to describe the training data set specifically.

\begin{figure*}[!htbp]
 \centering
\includegraphics[scale=0.48,keepaspectratio=true,clip=true,trim=0mm 0mm 0mm 0mm]{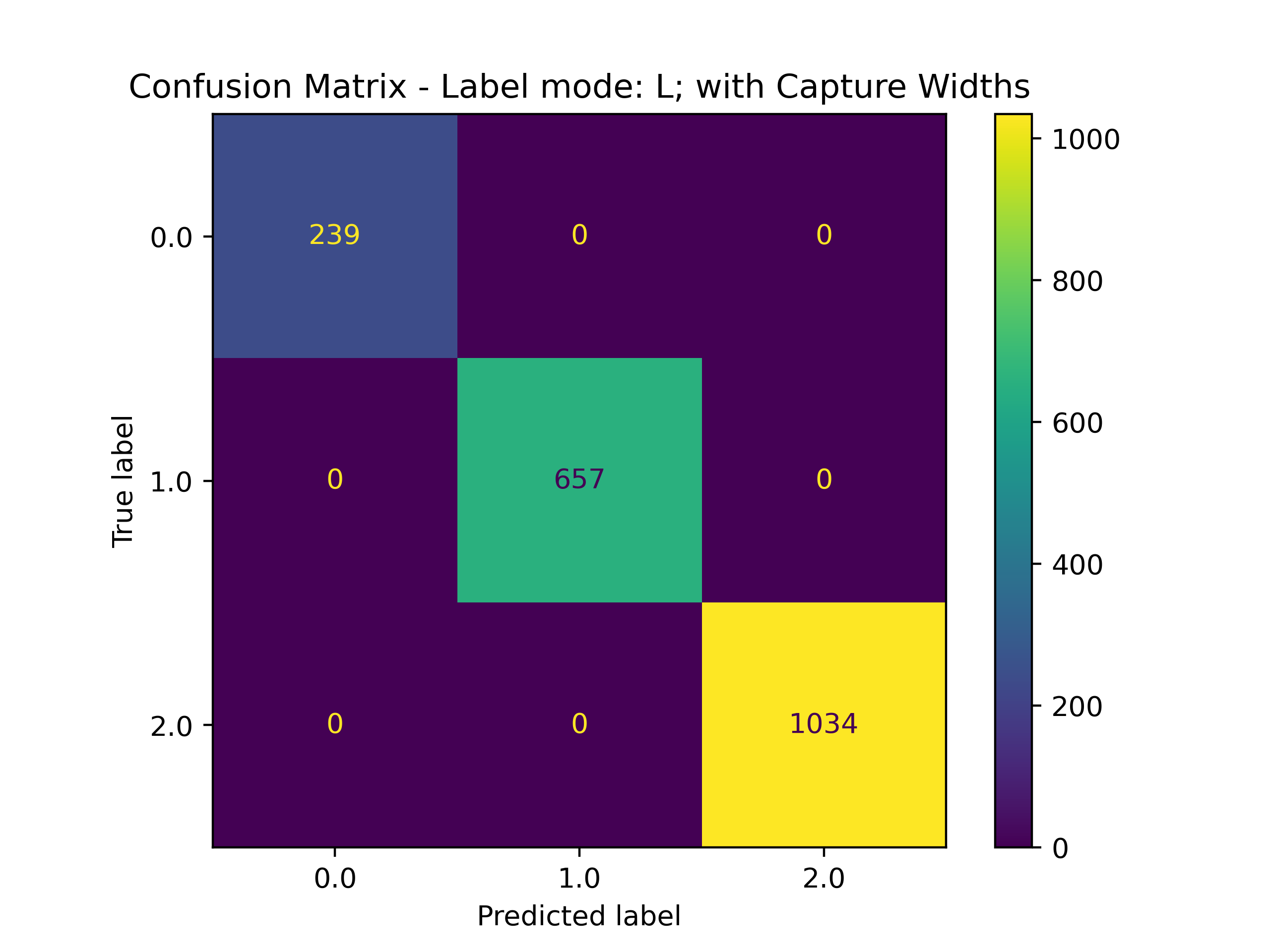}
\includegraphics[scale=0.48,keepaspectratio=true,clip=true,trim=0mm 0mm 0mm 0mm]{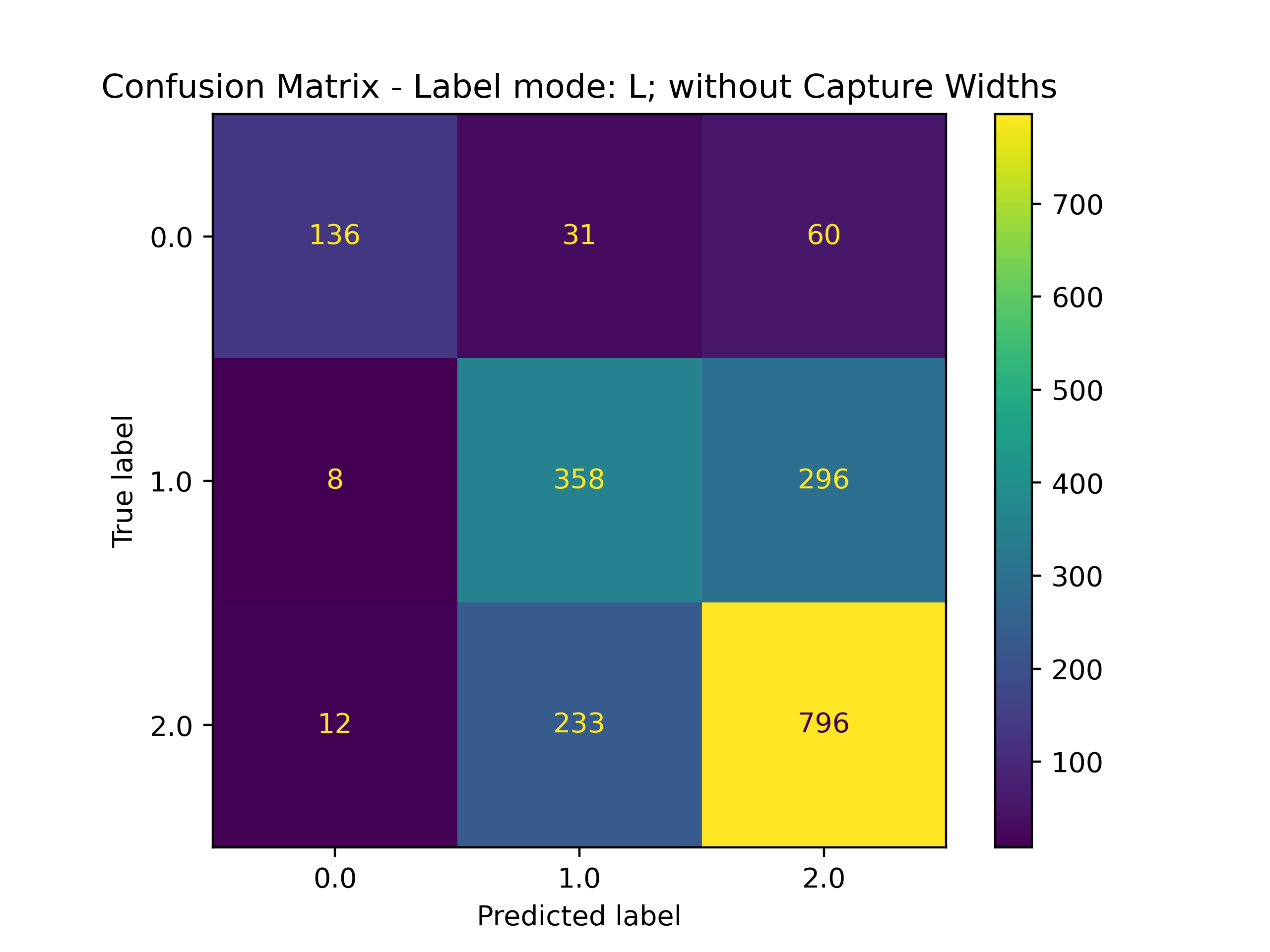}
\\
\includegraphics[scale=0.48,keepaspectratio=true,clip=true,trim=0mm 0mm 0mm 0mm]{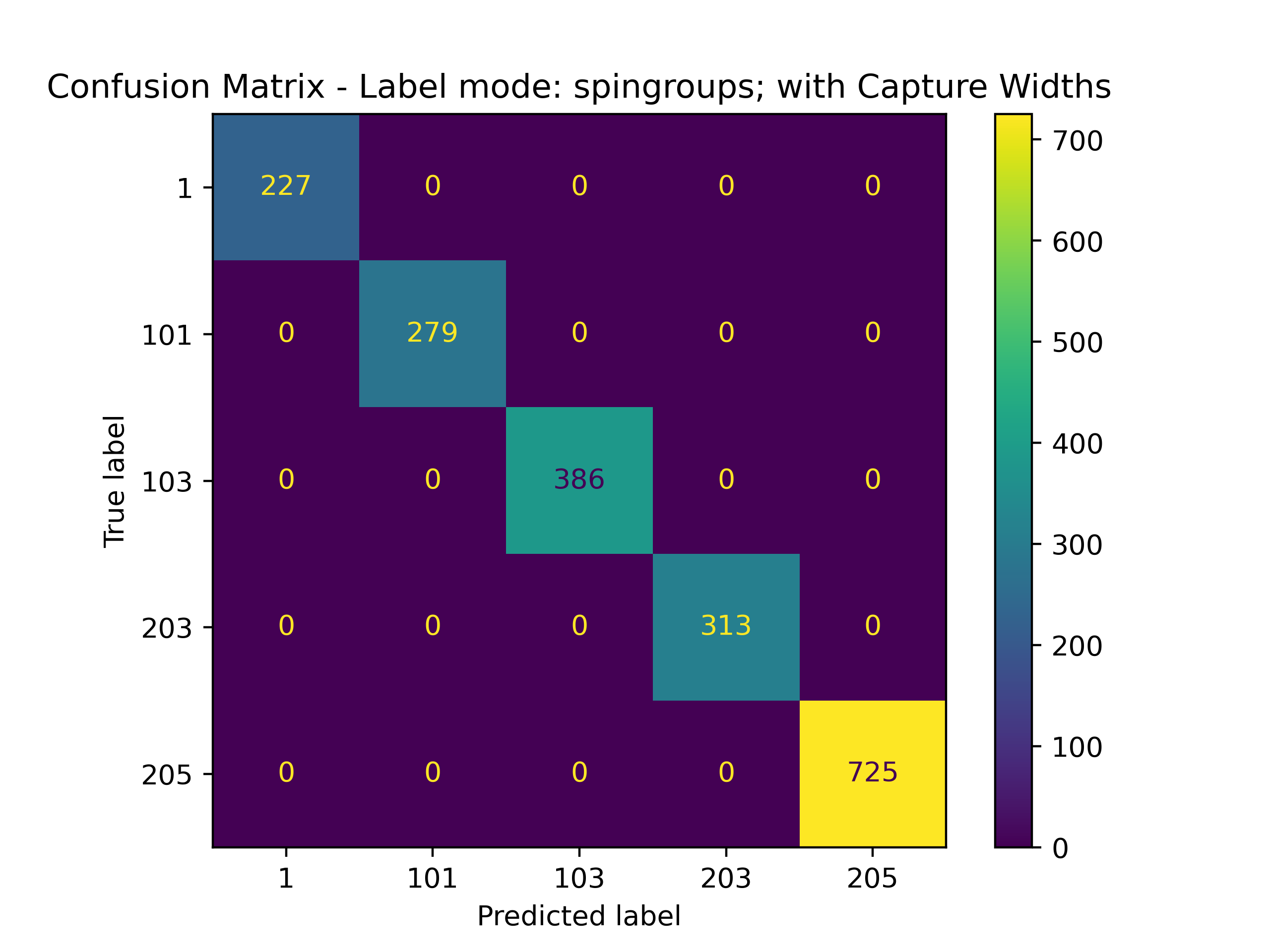}
\includegraphics[scale=0.48,keepaspectratio=true,clip=true,trim=0mm 0mm 0mm 0mm]{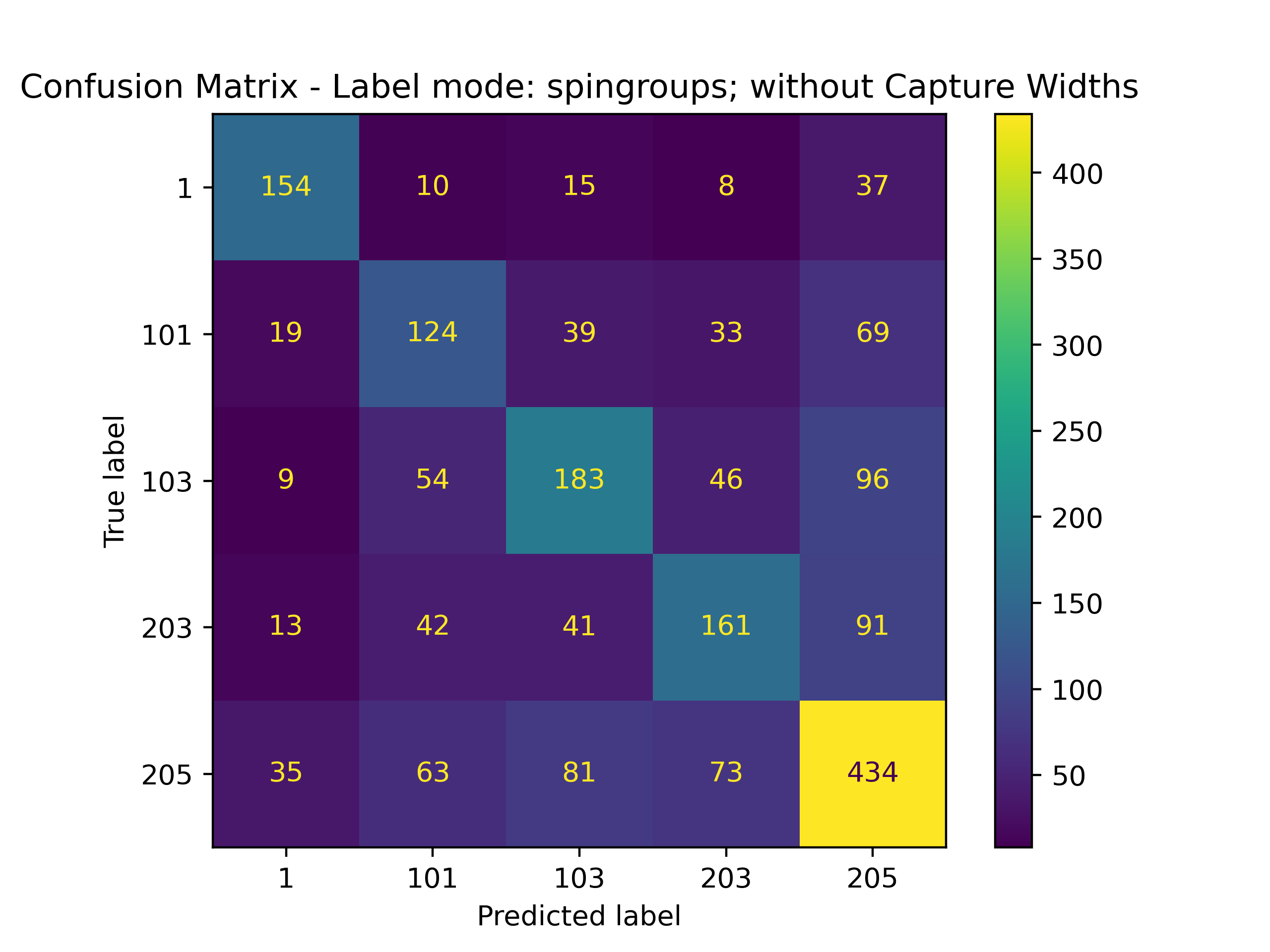}
\caption{Examples of confusion matrices obtained in the training process, taken from a single training event when training with a synthetic sequence with RMF=50\%. Each panel shows a different combination of label mode ($L$ and spingroup) and adoption or not of  features related to capture widths. Spingroup labels are described as $100*L+J$.}
\label{fig:confusion_matrices}
\end{figure*}

The training was performed both with and without the use of features that use the capture widths and categorizing either by $L$ or full spingroup.   Fig.~\ref{fig:confusion_matrices} shows examples of the typical confusion matrices that are obtained by the classifier in the training process, taken from a single training event when training with a synthetic sequence with RMF=50\%. We see the excellent training performance when capture widths are considered. However, as it will be further discussed in the text, this may be due to a strong training bias that may not translate to high-quality predictions if the trained classifier is applied to real resonance data.
Many of the aspects seen in Fig.~\ref{fig:confusion_matrices} are discussed in more detail later in the current work, where we consider results averaged over many training events and training sequences with different RMFs.

\begin{figure}[h]
\includegraphics[width=0.98\columnwidth,keepaspectratio=true,clip=true,trim=4mm 20mm 5mm 3mm]{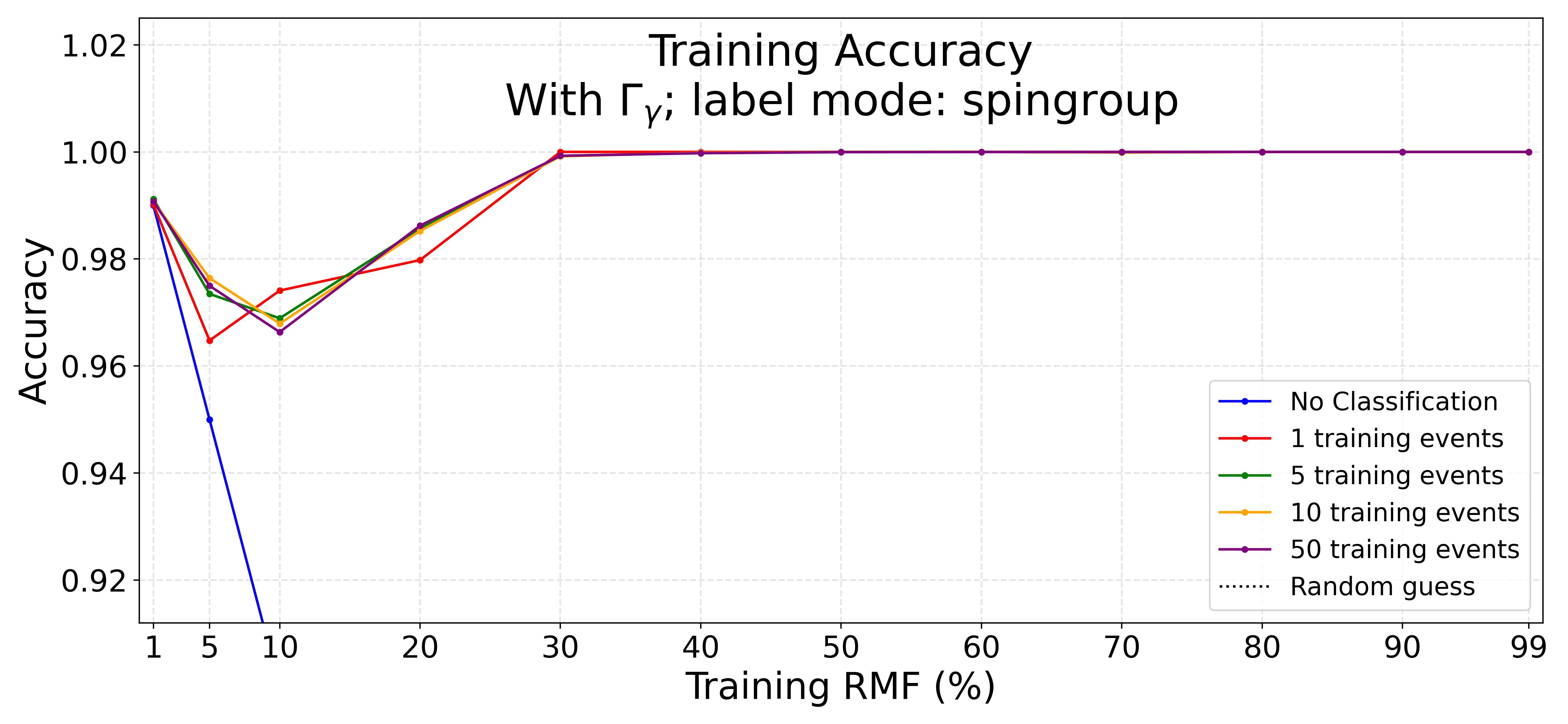}
\\
\includegraphics[width=0.98\columnwidth,keepaspectratio=true,clip=true,trim=4mm 20mm 5mm 3mm]{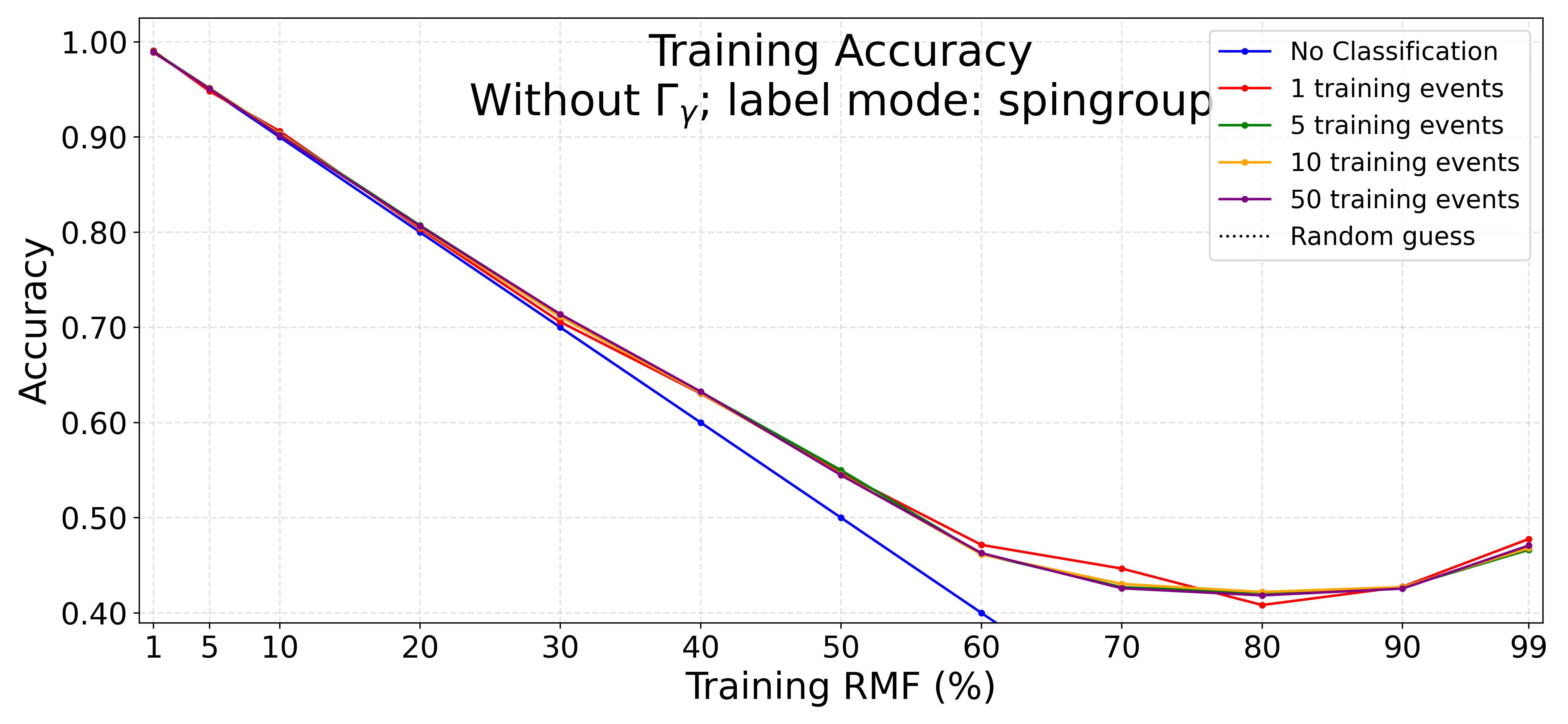}
\\
\includegraphics[width=0.98\columnwidth,keepaspectratio=true,clip=true,trim=4mm 20mm 5mm 3mm]{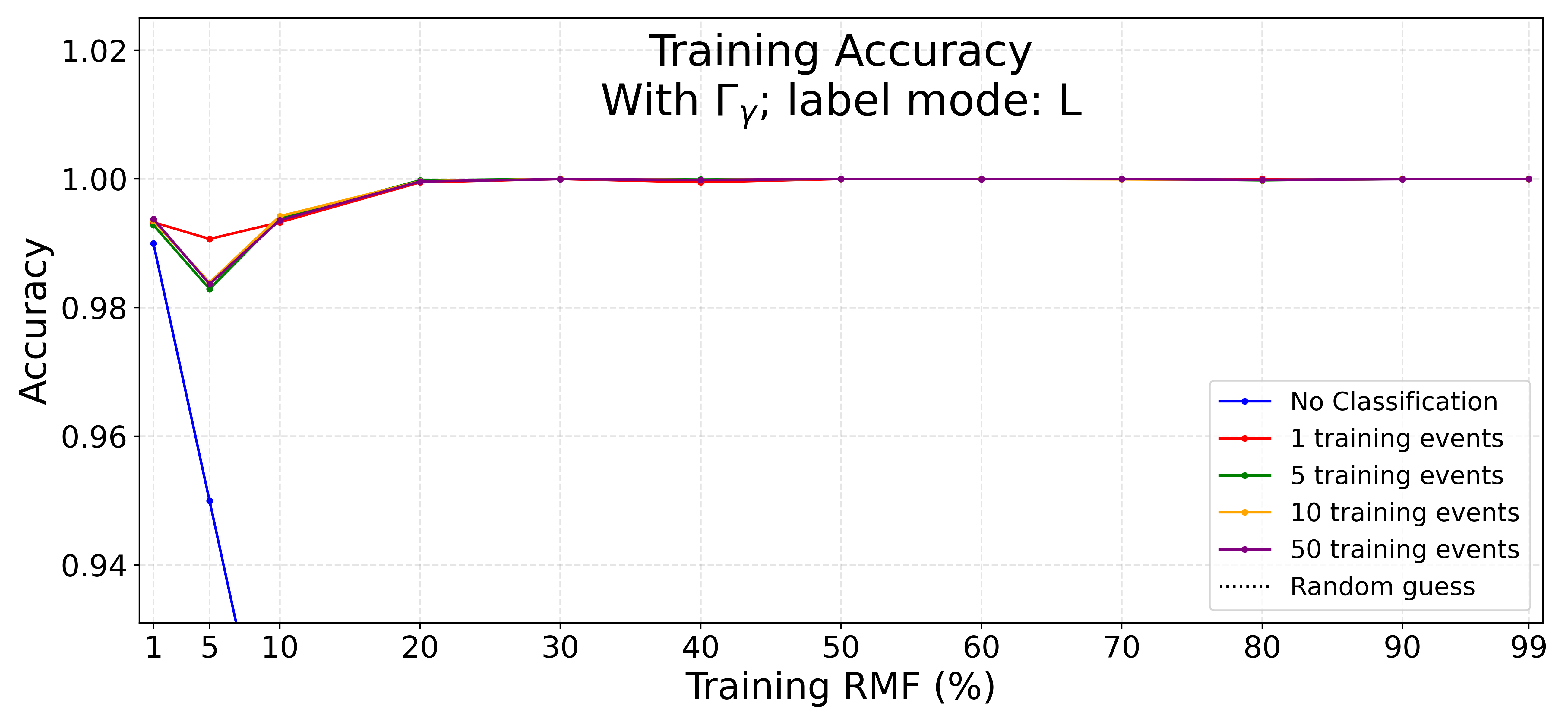}
\\
\includegraphics[width=0.98\columnwidth,keepaspectratio=true,clip=true,trim=4mm 0mm 5mm 3mm]{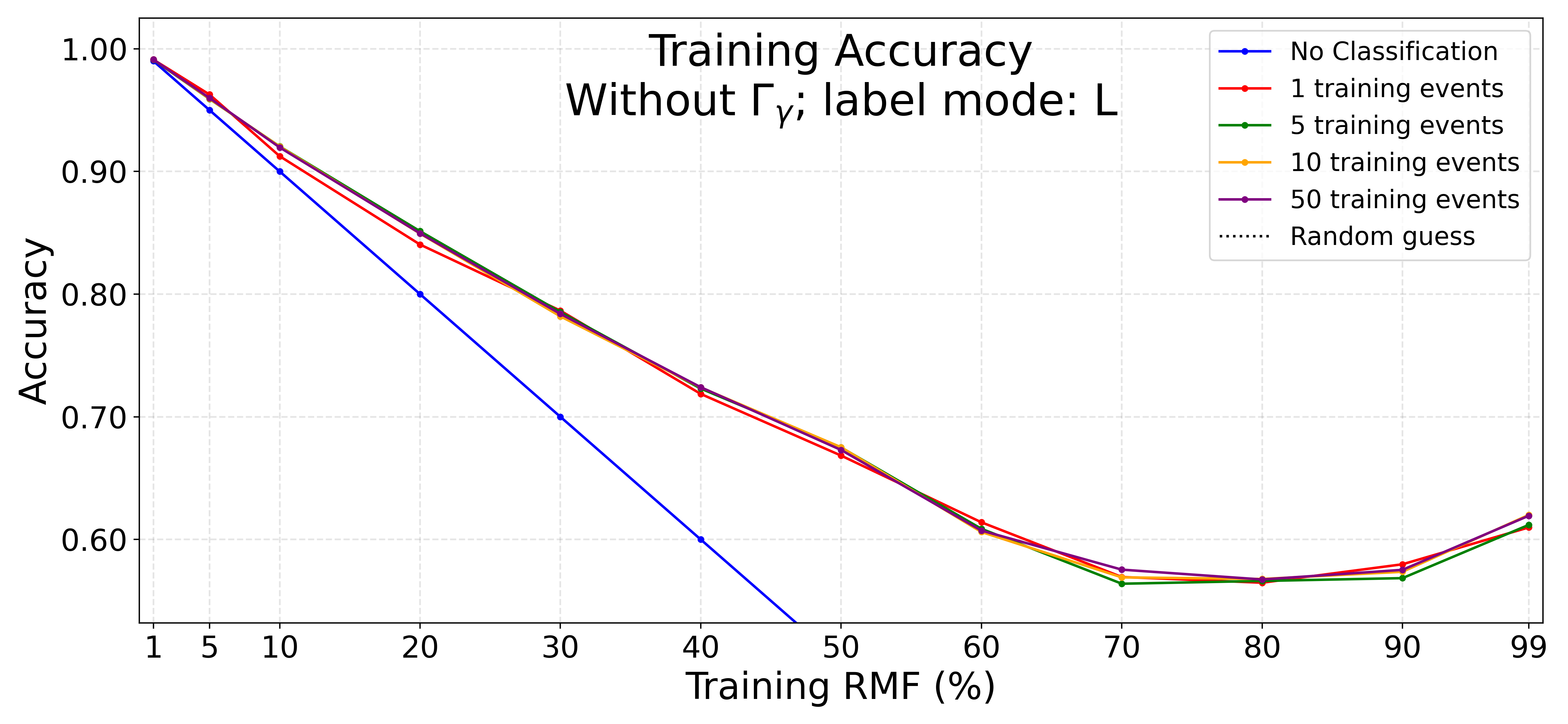}
\caption{Training accuracies for considering different number of training events, as a function of the RMF in the training set. Each panel shows a different combination of label mode ($L$ and spingroup) and adoption or not of  features related to capture widths. We show also a ``No classification'' curve corresponding to the original accuracy of the training set (1 minus the training RMF), which is the accuracy if no classification effort is made on that particular resonance sequence.  We also plot, although it is off-scale, the ``na\"{i}ve'' constant accuracy that one would get if choosing randomly among the allowed labels (1/3 for classification by $L$; 1/5 for classification by spingroups).}
\label{fig:training_accuracy}
\end{figure}

To quantify the performance of the classifier, we calculated accuracies based on the fraction of resonances that have the correct label. We are aware that there are many other important performance metrics (precision, recall, ROC curves, etc.)~\cite[chapter 3]{ML_Book} that would complement the accuracy analysis and help develop a full picture of the results and optimization pathways. However, being a work focused on the proof-of-principle of the method, we leave such more complete analysis for a future work.
Fig. \ref{fig:training_accuracy} shows the average training accuracy of the classifier as function of the misassigned fraction of the training set for all the combinations of label mode option (by $L$ or spingroup) and usage of capture width features. Each curve represents an average obtained with a different number of maximum training events, showing that by 50 training events the accuracy of each run has converged to the corresponding average accuracy.

When using capture widths, whether classifying by $L$ alone or by spingroup, we achieve nearly perfect reclassification. However, even though capture widths can be very discriminative, this may not be a reasonable feature option when applying to the classification of real resonance distributions that may contain biases towards average widths, as discussed in Section~\ref{paragraph:capwidth}.  Indeed, because we chose $\nu_\gamma\rightarrow\infty$, our capture width distributions are essentially delta functions so a perfect capture width match is needed to be considered in the distribution for a given label. 

When capture width features are not employed, we see a consistent pattern of highest training accuracy for low-misassigned datasets, with average accuracies decreasing as misassignment increases until it flattens or even upends with mostly misclassified sets. As a reference, each plot in Fig. \ref{fig:training_accuracy} shows the ``do nothing'' line -- a line representing the degree of accuracy of the training set (basically a line function 1 - RMF), which indicates the overall accuracy of the set if no classification attempt is made. We also plot reference lines corresponding to ``random guess,'' being the accuracy one obtains by randomly selecting among the allowed labels. The accuracies obtained, though, are much higher than the random guess, so the lines are off the scale in  Fig. \ref{fig:training_accuracy}.  

For classification by $L$, the training accuracy decreases from about 99\% accuracy at low  RMF to a minimum of $\sim$60\% average accuracy at 70-80\% RMF. It is noteworthy, however, that this is still much more accurate than the ``no classification'' baseline. On the other hand, when classifying by spingroup, the classifier has more difficulty to  assign the correct label, with average training accuracies remaining closer to the ``no classification'' line at low RMFs and stabilizing at  $\sim$40\% for higher RMFs, although still much higher than simply guessing. This is somewhat expected as there are five possible labels when classifying by spingroups instead of only three with label mode $L$, making it a more difficult problem to solve.

\begin{figure}[!htbp]
\includegraphics[width=0.98\columnwidth,keepaspectratio=true,clip=true,trim=4mm 20mm 5mm 3mm]{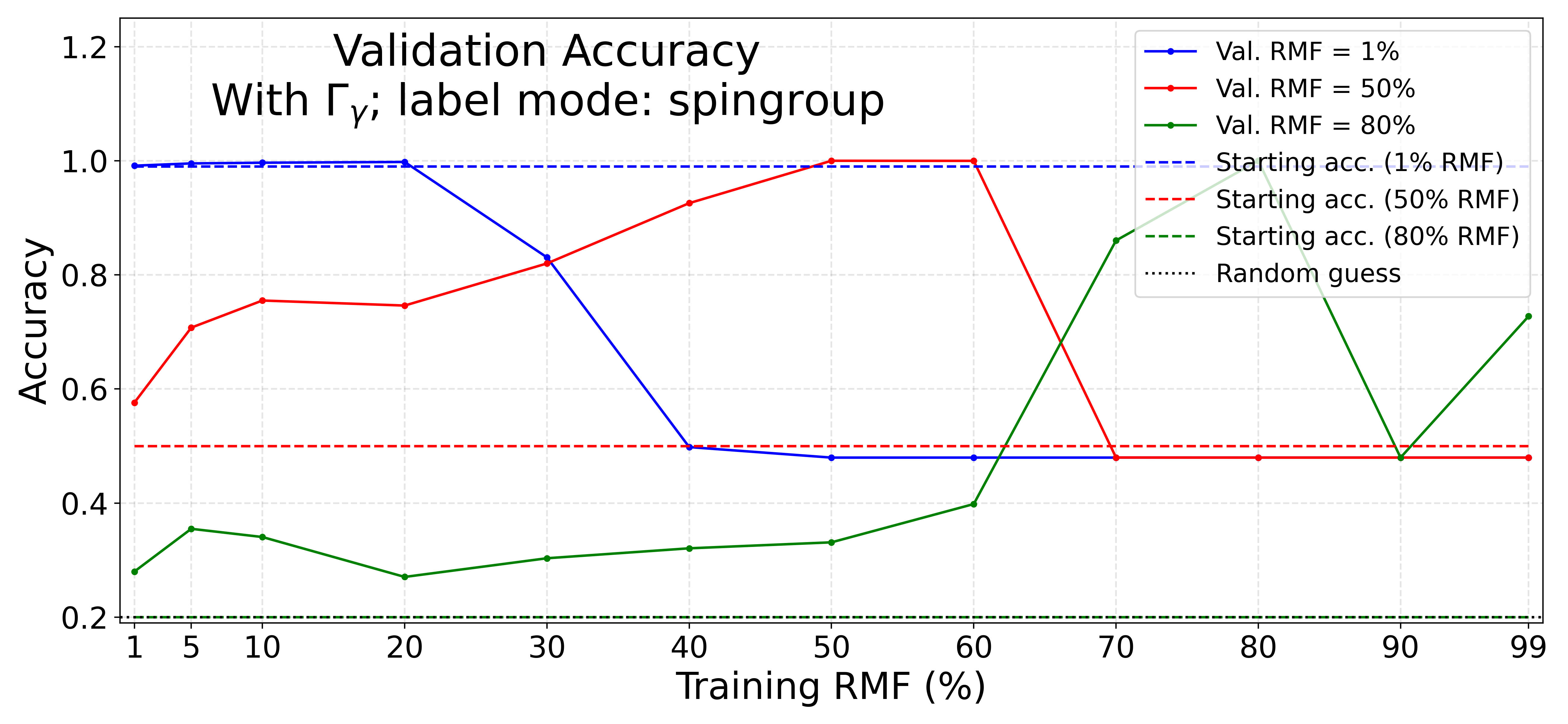}
\\
\includegraphics[width=0.98\columnwidth,keepaspectratio=true,clip=true,trim=4mm 20mm 5mm 3mm]{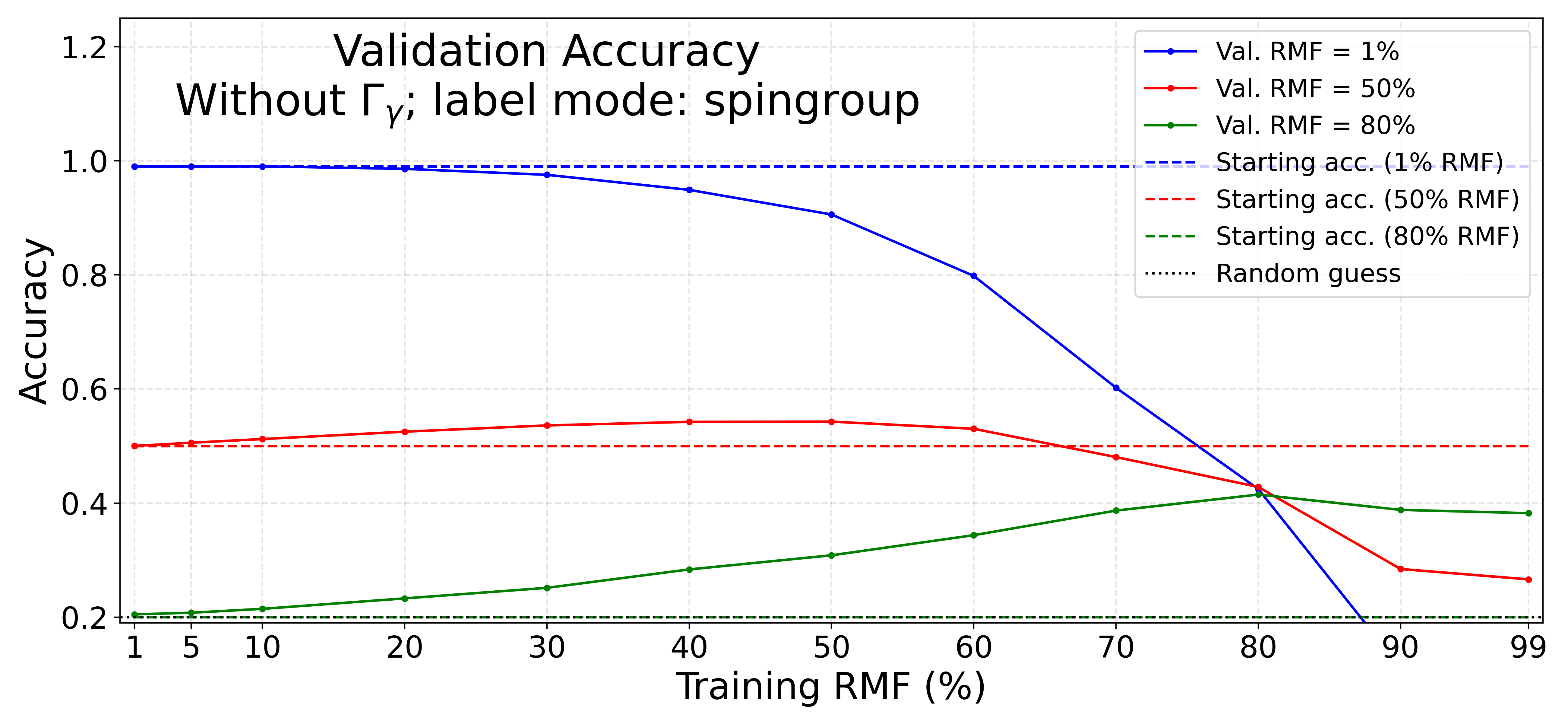}
\\
\includegraphics[width=0.98\columnwidth,keepaspectratio=true,clip=true,trim=4mm 20mm 5mm 3mm]{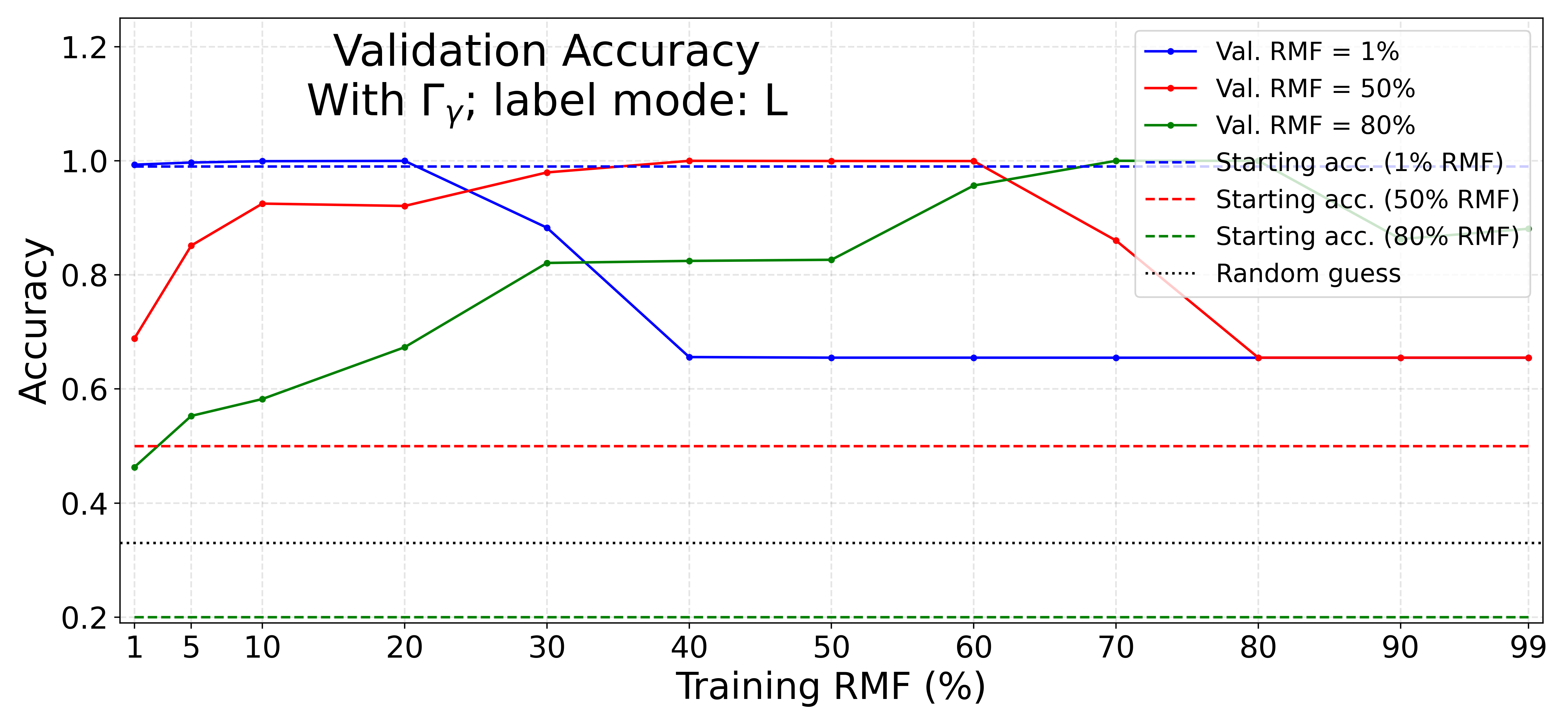}
\\
\includegraphics[width=0.98\columnwidth,keepaspectratio=true,clip=true,trim=4mm 0mm 5mm 3mm]{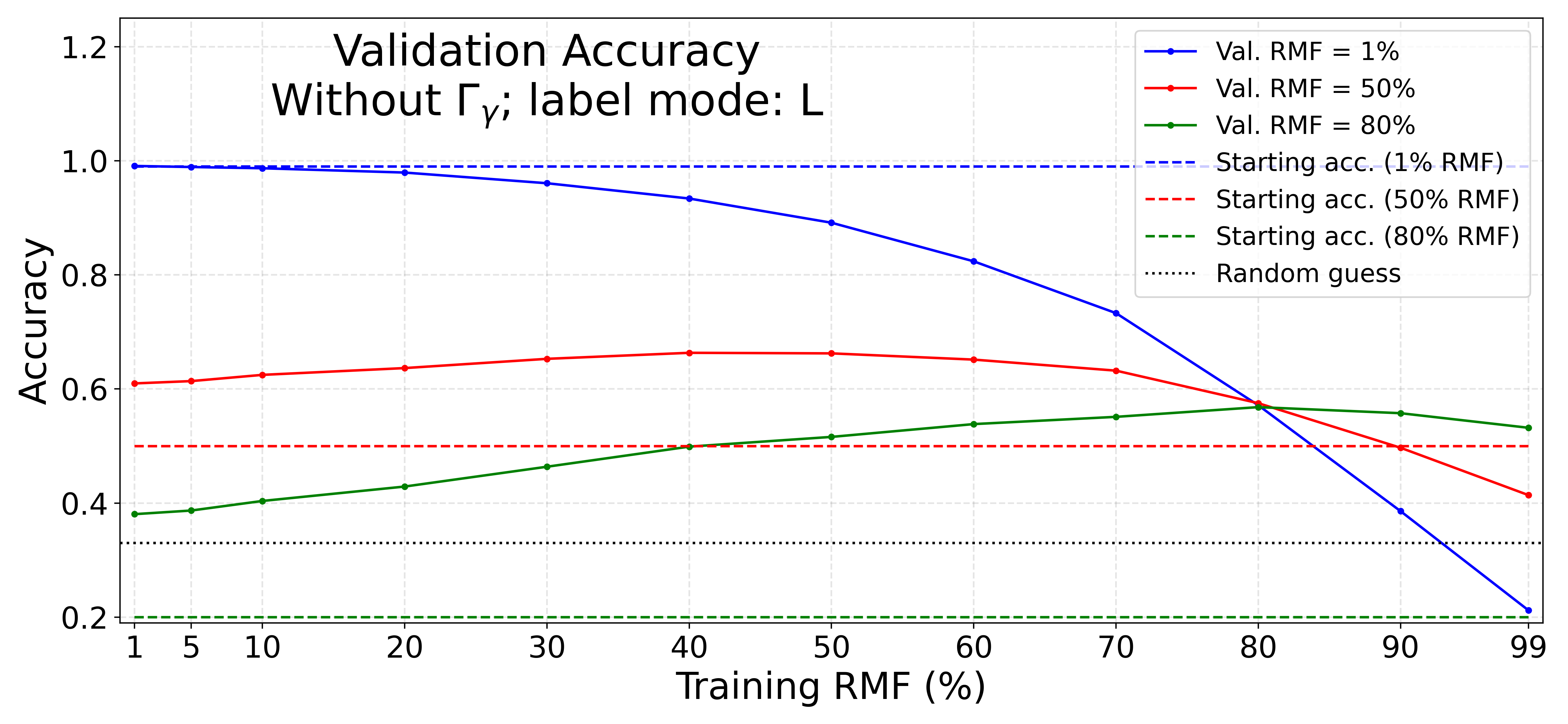}
\caption{Validation accuracies for different RMFs in the synthetic validation set, as a function of the RMF in the training set (solid lines). Each panel shows a different combination of label mode ($L$ and spingroup) and adoption or not of  features related to capture widths. We show as dashed lines the original accuracy of the validation set (1 minus the validation RMF), which is the accuracy if no classification effort is made on that particular resonance sequence, with the same color of the validation accuracy for the corresponding sequence.  We also plot, although it is sometimes off-scale, the ``na\"{i}ve'' constant accuracy that one would get if choosing randomly among the allowed labels (1/3 for classification by $L$; 1/5 for classification by spingroups).}
\label{fig:validation_accuracy}
\end{figure}

\subsection{Validating on synthetic data}
\label{sec:validation_synthetic}

Once the performance of the classifier during the training process was better understood and benchmarked, we validated the method by applying the fully trained MLP algorithm to a second realization of synthetic data based off \nuc{52}{Cr}. We again implemented random misassignments to this second realization of synthetic data with RMFs ranging from 1\% to 99\%. Fig.~\ref{fig:validation_accuracy} shows representative results of the validation analysis.

In Fig.~\ref{fig:validation_accuracy}, we show the validation accuracies averaged over 50 training events for the validation sequences having RMFs of 1\%, 50\%, and 80\%, as a function of the RMFs in the training set, represented by the solid lines. We also display as dashed lines of corresponding colors the starting accuracies of each validation sequence (e.g., the validation sequence with 80\% RMF is 20\% correct). A comparison from the dashed line with the solid line of the same color shows how much the machine-learning classification has improved (or worsened) the set relative to the original resonance sequence. In all cases, we see that the maximum  validation accuracies happen when the training sequences have  around the same RMFs as the sequences being validated. This is somewhat expected as those are the cases in which the validation sequences are the most statistically similar to the training sequences. Interestingly though, for classifications both by $L$ and by spingroup, these peaks in accuracy are much sharper when employing capture width features and much smoother when not using them. This indicates that capture widths are very discriminative. 
However, given the known bias in the use of the capture widths, 
we focus our discussion in the cases where capture widths were not used as a feature.

Firstly, we shall focus on the case of label mode $L$ without capture widths from Fig.~\ref{fig:validation_accuracy}, bottom panel. In the case of validation sequence with RMF of 1\% (blue line), the original sequence was already very accurate and for low RMFs in the training set the classifier preserves that, worsening it minimally with training sequences up to around 20\% RMF. Above that, the reclassification accuracy decreases quickly. For a validation sequence of RMF=50\% (red lines), the reclassified sequence is consistently more accurate than the original one, up to training RMF of around 90\%. For the validation sequence with RMF=80\%, the machine-learning algorithm provides a substantially more accurate sequence regardless of how much is the RMF for the training set. This shows that, with the appropriate training set (or range of training sets), the classifier is able to deliver a resonance sequence that is more accurate than the one provided as input. This suggests that an iterative process in which, under the appropriate conditions, a sequence of arbitrarily low  accuracy related to  $L$ assignments could be incrementally improved until being fully correct. The development of such iterative method will be pursued in a future work.

We now turn to the validation results of Fig.~\ref{fig:validation_accuracy}, second panel from top,  corresponding to label mode by spingroup, without capture widths. In this case, similar considerations can be made when the validation sequence initial accuracy is low (meaning high RMF), as is the case of the solid green curve corresponding to RMF=80\%. We see that the reclassified accuracy is consistently better than the original accuracy (dashed green curve), for all values of RMF in the training set. For lower validation RMFs (solid red and blue curves),  the accuracies as function of training RMF are similar to the case of label mode $L$, although a little lower. Also, resulting average accuracies seem closer to or lower than the initial accuracies (corresponding dashed lines) in a larger training RMF range, indicating that an iterative process may be trickier for spingroup classification than it would be for label mode $L$.  This may be explained by the fact that the classification by spingroup is much more challenging than by $L$: the number of possible labels is larger as for each $L \neq 0$ as there will be two spingroups allowed per $L$. Still, an iterative method for spingroups may still be  effective if one tackles it in two steps: first classifying by $L$, and later by spingroup within fixed $L$ values. Again, this is outside of the scope of this work and will be investigated in the future.

\subsection{Reclassifying real resonance data}

After validating the reclassification method in synthetic data with known RMF, 
we applied the trained algorithm to the ENDF/B-VIII.0 \nuc{52}{Cr} resonance data from Ref.~\cite{Nobre_2021}. 

\begin{figure*}[!htbp]
\includegraphics[scale=0.50,keepaspectratio=true,clip=true,trim=0mm 0mm 0mm 0mm]{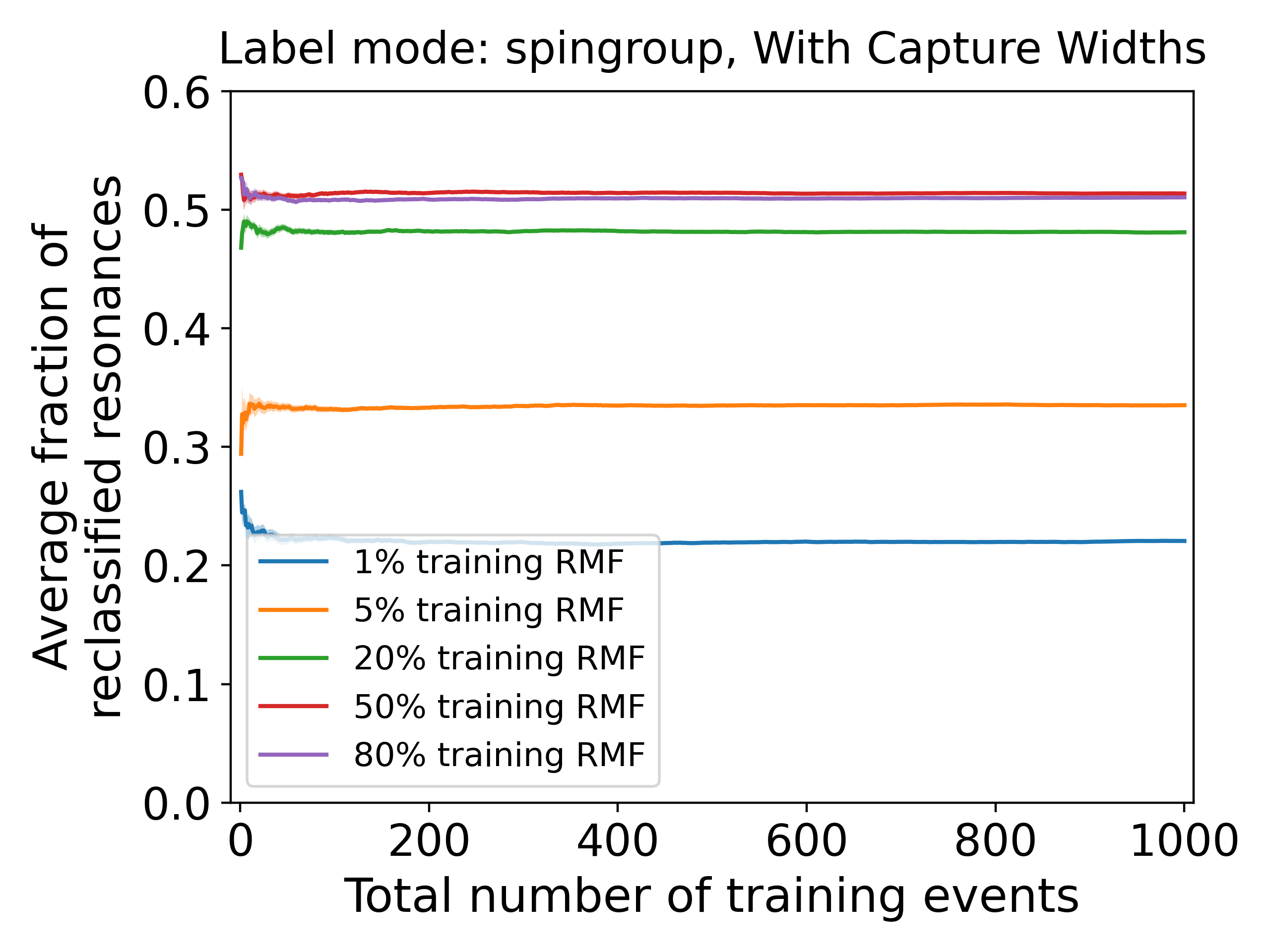}
\includegraphics[scale=0.50,keepaspectratio=true,clip=true,trim=0mm 0mm 0mm 0mm]{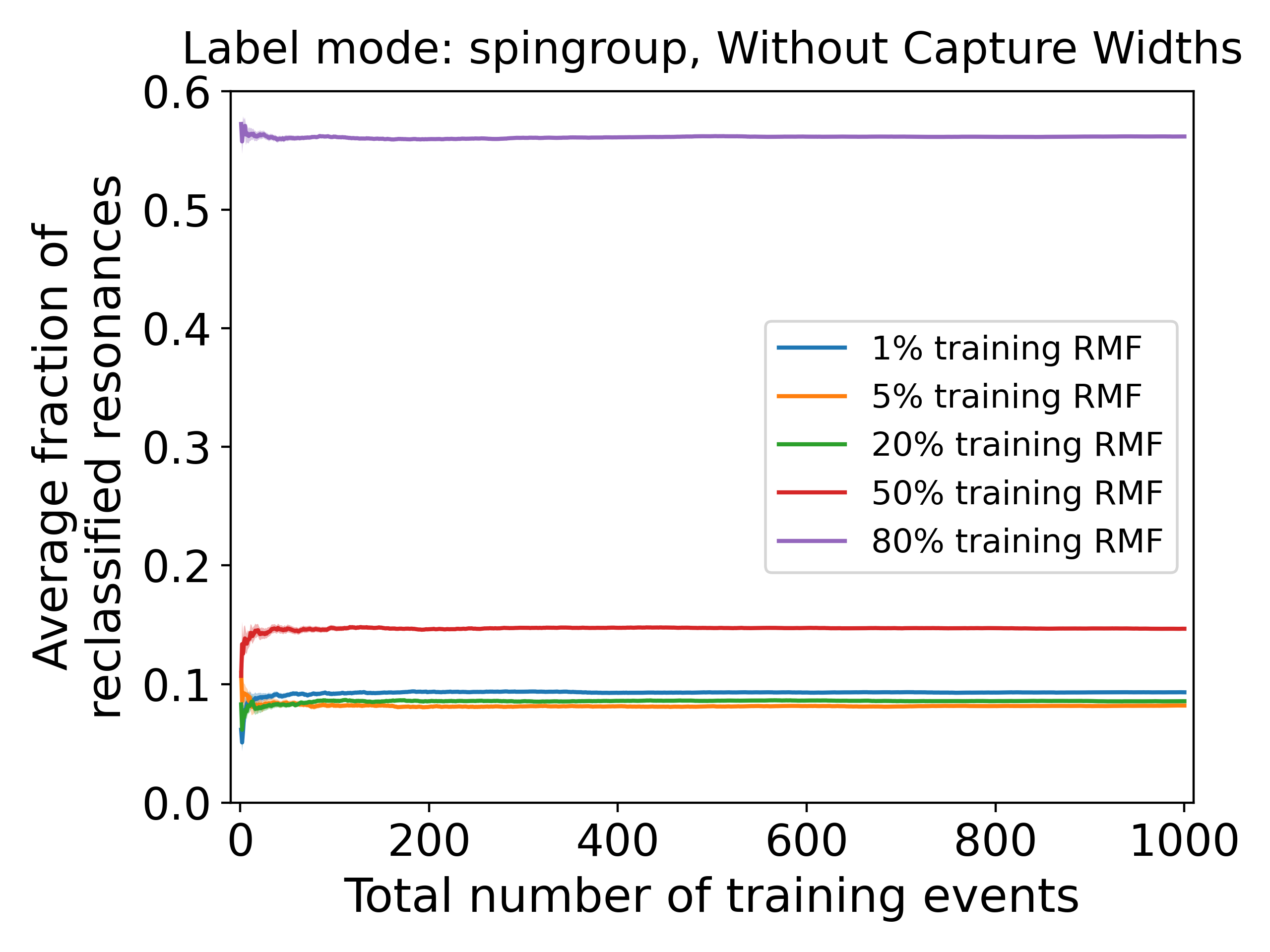}
\includegraphics[scale=0.50,keepaspectratio=true,clip=true,trim=0mm 0mm 0mm 0mm]{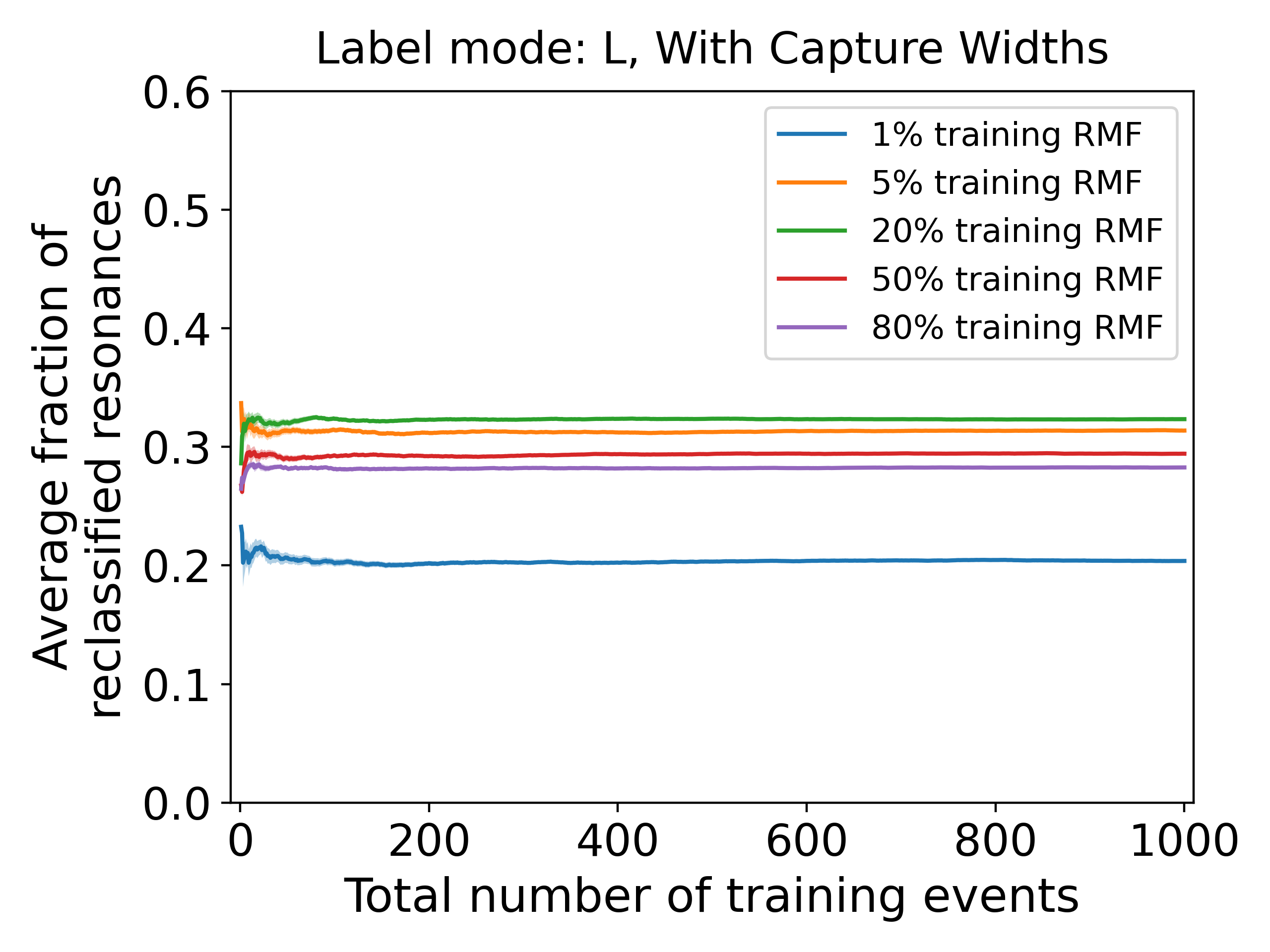}
\includegraphics[scale=0.50,keepaspectratio=true,clip=true,trim=0mm 0mm 0mm 0mm]{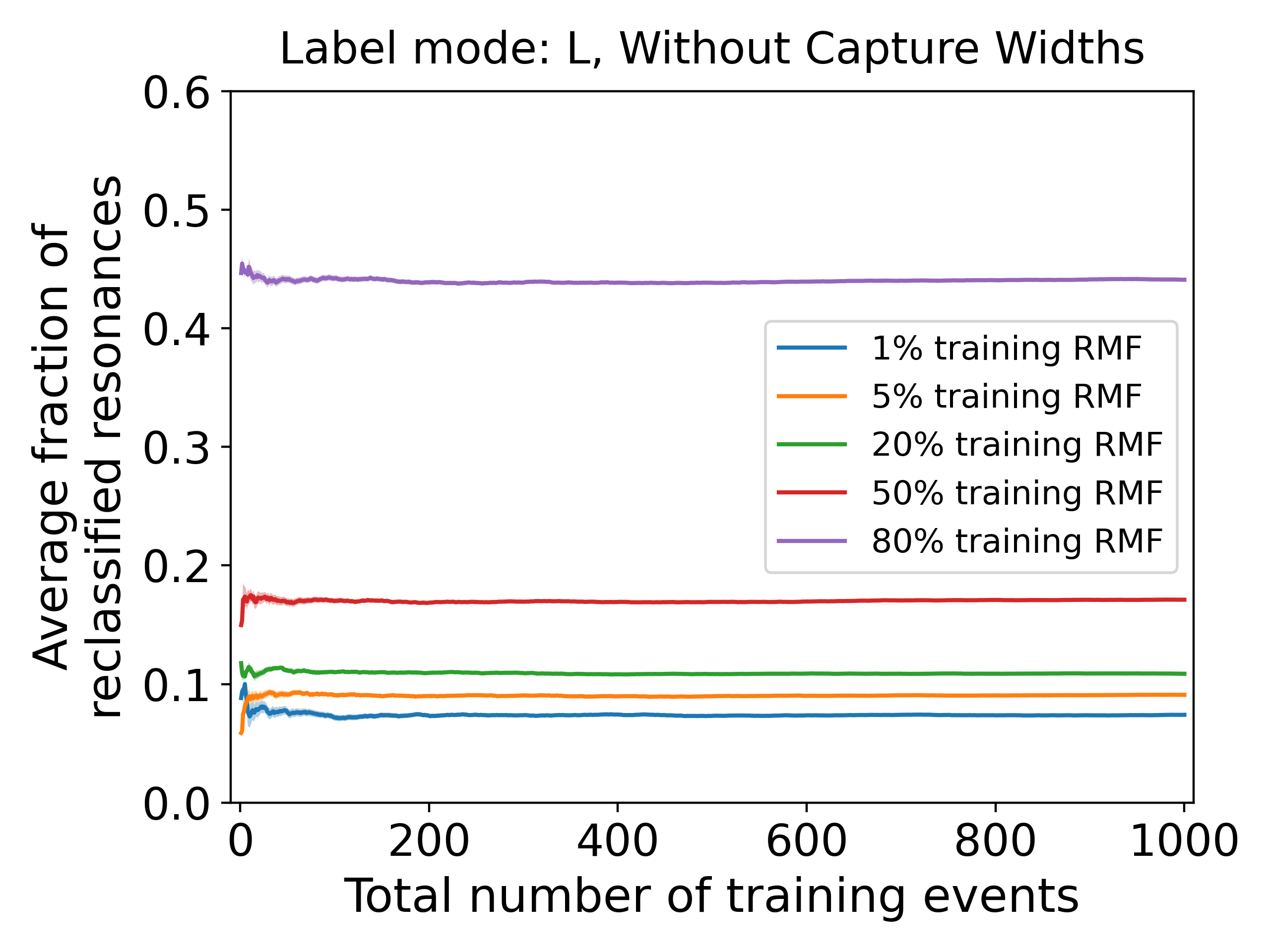}
\caption{Average fraction of ENDF resonances reclassified as a function of total number of training events. Although the average values oscillate with a small number of training events, they converge well after around 200-400 training events.}
\label{fig:fraction_reclassifed}
\end{figure*}

The first step is to estimate how many training events are needed to allow us to assume the reclassification process has converged. 
For that, we determined the average fraction of evaluated resonances that were reclassified as a function of the maximum number of training events considered as shown in Fig.~\ref{fig:fraction_reclassifed}.  
Here we show the resulting fraction of reclassified ENDF resonances for different values of RMF in the training set. 

We see that by 1000 training events, all values of fraction of reclassified resonances have clearly converged to their average value. 
As a matter of fact, for all cases the average fraction of reclassified resonances converge after around 200-400 training events.  
While for label mode $L$ without capture widths,  the average fraction of reclassified resonances seem to always increase as the training RMF increases; this is not an observed trend in the other cases.

\begin{figure*}[!htbp]
 \centering
\includegraphics[scale=0.52,keepaspectratio=true,clip=true,trim=0mm 0mm 0mm 0mm]{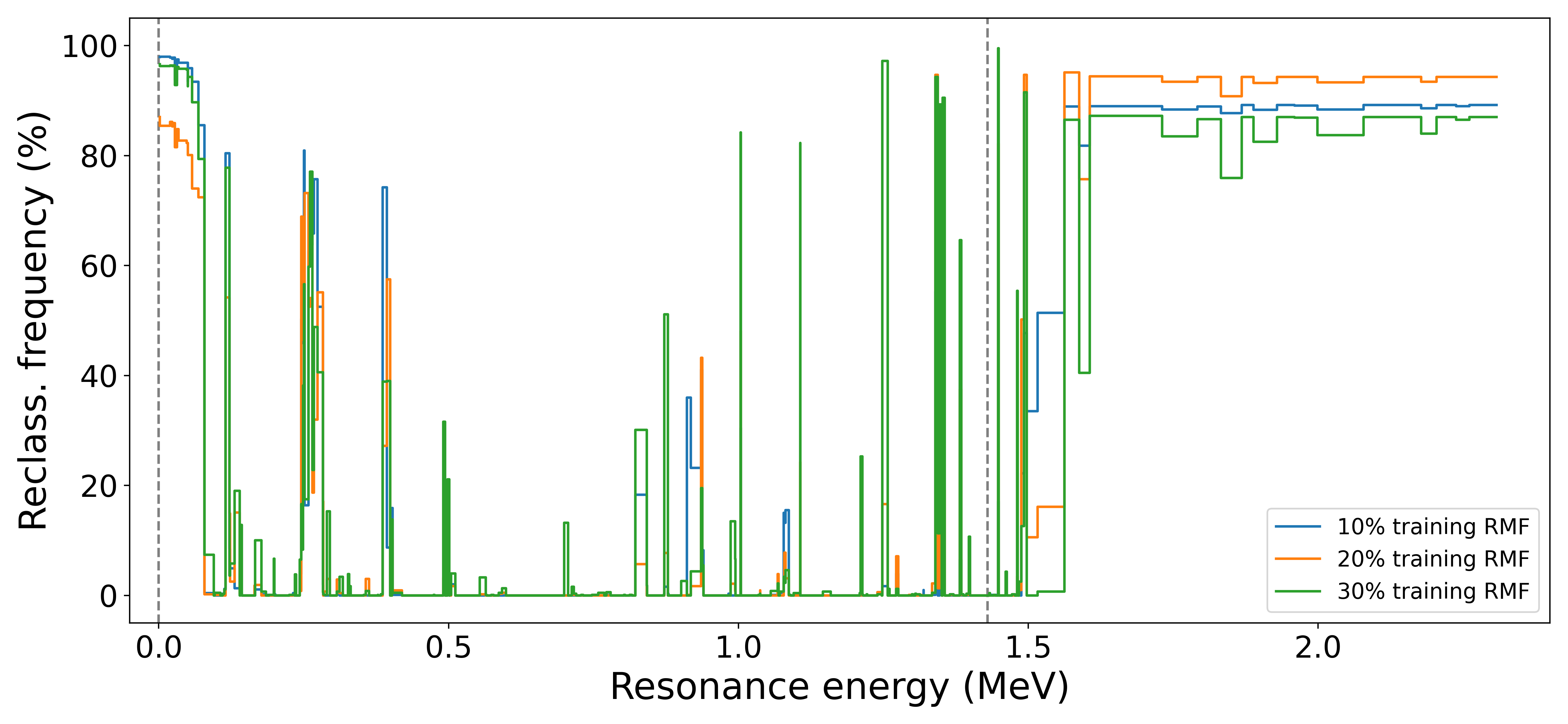}
\caption{Frequency of reclassification of each ENDF resonance. The dashed-gray vertical lines indicate the beginning and end of the resonance region in the evaluated file.}
\label{fig:fraction_reclassifed_real}
\end{figure*}

We turn our attention for the individual resonances from the evaluated file that are being reclassified. 
From the discussion above, it is clear that we cannot trust results using the capture width distribution.  
Further, from the discussions in Section~\ref{sec:validation_synthetic}, we see that the most reliable reclassification process is obtained by classifying only by $L$.
With these, we require a training set that has a RMF that is similar to the one being reclassified. 
However, it is challenging to define \emph{a priori} what is the real RMF of a resonance sequence in an ENDF-evaluated file that originates from real measured data. 
To proceed, some realistic considerations based on expert judgement is necessary. 
It is very unlikely that the resonances for the major isotope of a well-known, well-measured material, such as chromium would have more wrong spin assignments than correct ones. 
At the same time, it is unrealistic to assume that practically all assignments are correct. 
It is thus reasonable to assume that the RMF in real data of \nuc{52}{Cr} is somewhere in the range between $\sim$ 10\% to 50\%. 
From Fig.~\ref{fig:fraction_reclassifed}, for the cases without capture widths, we see that the fraction of reclassified resonances does not change much around training RMF=20\%, with RMF=50\% beginning to distance from lower RMFs, indicating that the reclassification process for the evaluated resonance data is somewhat stable at RMF=20\%. 
For this reason, we show in Fig.~\ref{fig:fraction_reclassifed_real} the normalized number of times each ENDF resonance was reclassified by the MLP algorithm trained on synthetic data with 20\% RMF over the course of 1000 training events, as a function of the resonance energy. 
As a stability test, we also plot the results using training sets with RMF=10\% and 30\%. 

\begin{figure*}[!htbp]
 \centering
\includegraphics[scale=0.52,keepaspectratio=true,clip=true,trim=0mm 0mm 0mm 0mm]{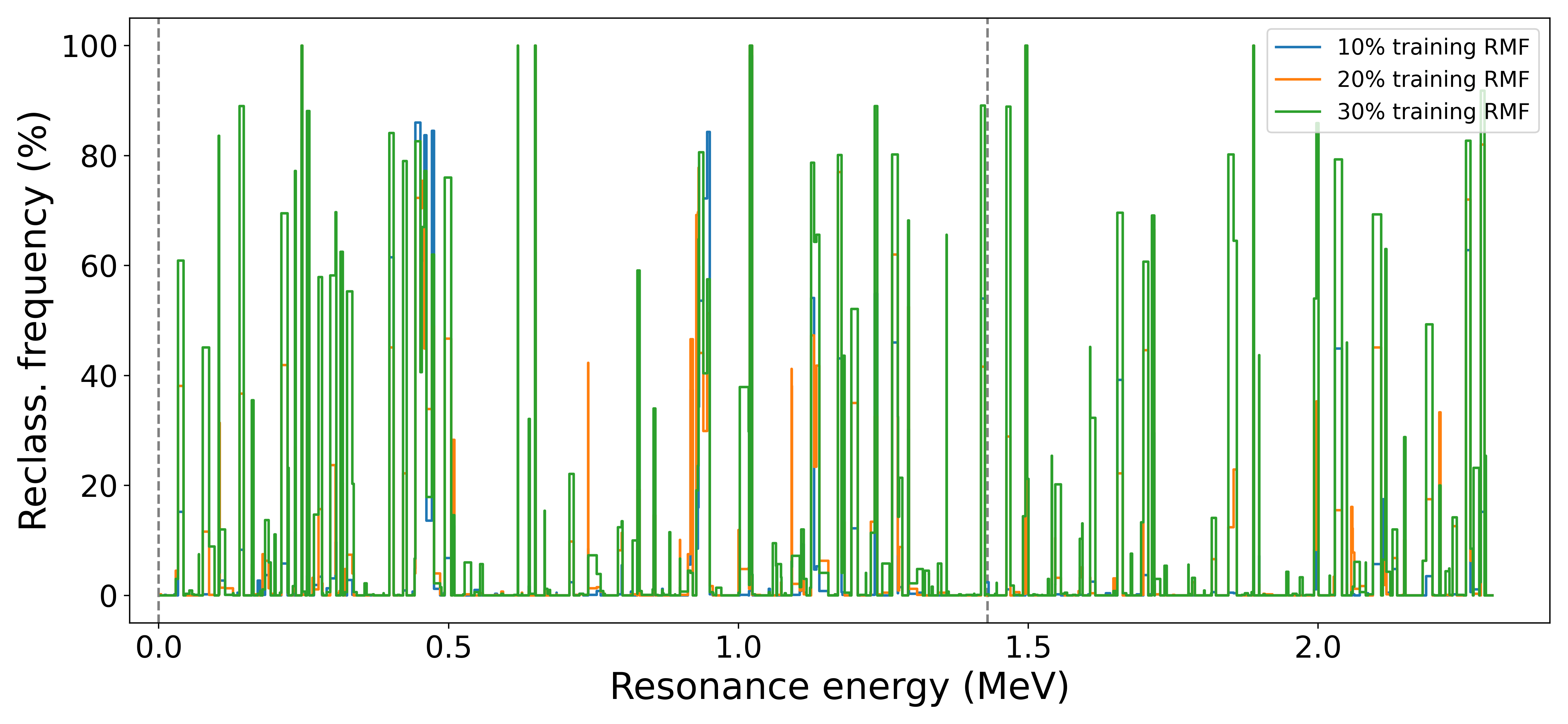}
\caption{Frequency of reclassification of each  resonance in a synthetic sequence with 20\% RMF.    The dashed-gray vertical lines indicate the beginning and end of the resonance region in the evaluated file.}
\label{fig:fraction_reclassifed_synth}
\end{figure*}

We see in Fig.~\ref{fig:fraction_reclassifed_real} that indeed there is very little difference among the calculations with training data of the different RMFs listed. In general, we observe many regions in which no resonances are reclassified, or some of them very rarely. There are, on the other hand, some resonances, and sometimes, cluster of resonances, that are frequently, if not almost always reclassified.  In particular, we note the two clusters of reclassified resonances: one near the beginning of the sequence and the other at the end, above $\sim$1.6 MeV. To rule out any intrinsic bias from this classification process, we repeated the exact same calculations, but this time, instead of applying the trained algorithm into real data, we applied it to an independent realization of synthetic data with 20\% RMF. This is shown in Fig.~\ref{fig:fraction_reclassifed_synth}. We see that the peaks of resonances reclassified most times for the synthetic sequence seen in Fig.~\ref{fig:fraction_reclassifed_synth} are more randomly distributed, without significant clusters. This is expected since the synthetic sequence had 20\% of its resonances missasigned randomly. This lends confidence that the real resonances reclassified in multiple training events, with multiple training seeds, seen in  Fig.~\ref{fig:fraction_reclassifed_real} may actually correspond to incorrect assignments.

\begin{figure*}[!htbp]
 \centering
\includegraphics[scale=0.64,keepaspectratio=true,clip=true,trim=25mm 3mm 31mm 15mm]{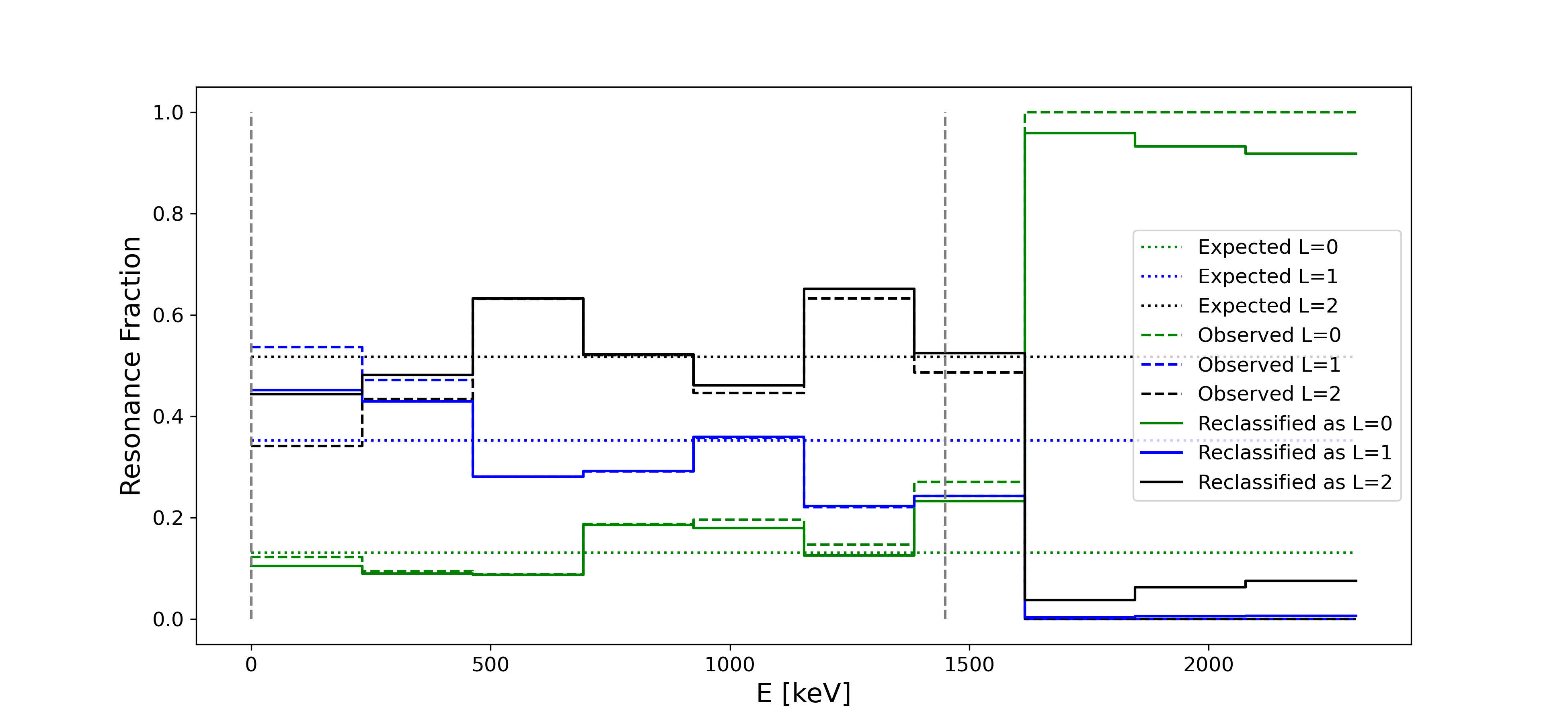}
\caption{Plot showing the observed and reclassified fraction of resonances in a given energy interval.  The expected fraction of resonances is taken using all resonances in the ENDF/B-VIII.0 file.  For a given $L$, over a given energy interval $\Delta E$, the expected number of resonances is given by $N_L=\Delta E/D_L$; therefore, the fraction of resonances of a given $L$ is $N_L/N_{tot}=D/D_L$ with $D$'s taken from Table \ref{table:52Crparams_real}.  The dashed vertical lines denote the limits of the resonance region in the ENDF file.}
\label{fig:fraction_reclassified_real_by_L}
\end{figure*}

It is instructive to deconstruct the results shown in Fig. \ref{fig:fraction_reclassifed_real} by orbital angular momentum in order to see if there are correlations in the resonances the reclassified resonance assignments.  This is shown in Fig. \ref{fig:fraction_reclassified_real_by_L}, broken into 10 equally spaced energy groups.  We see that the resonances above 1450 keV are originally assigned to $L=0$ and the reclassifier is attempting to reclassify them mainly to $L=2$.  
In this evaluated set of resonances, the resonances above 1450 keV were added to provide a background and are not expected to be correctly classified.
Interestingly, we see a similar behavior of the reclassifier in the lowest energy group. However, instead of reclassifying the $L=0$ resonances, it is reclassifying the $L=1$ resonances to $L=2$.   It is clear that the classifier expects more $L=2$ resonances than are observed in the evaluation.  What is less clear is whether we should trust the classifier's assignments any more than the original evaluator's expert judgement.

To further explore the classifier's choices, we show which were the real resonances that were reclassified more than 50\% of the time and the distribution probability of the reclassification label in Table~\ref{table:reclassENDF}.  Here only the 44 most reclassified resonances are listed which corresponds to  about $\sim$12\% of the total number of real resonances in the evaluated file.  This fraction is consistent with the asymptotic converged value for the average fraction of reclassified resonances for label mode $L$, without capture width features, and 20\% training RMF, as seen in Fig.~\ref{fig:fraction_reclassifed}.  In Table \ref{table:reclassENDF},  we also show the $L$ assignments from other resonance quantum number determinations in the literature.  These references and the methods used to make their quantum number determinations are given in Table \ref{table:52Cr_references}.  Interestingly, the 25 most commonly reclassified resonances were not observed by any of the authors in Table \ref{table:52Cr_references} and the $L$ determination is based solely on the shape analysis of Leal \etal \cite{ND2010_Cr}.   If we were to adopt the reclassifier's assignments over those in Ref.~\cite{ND2010_Cr}, it would not have much measurable impact on the reconstructed cross section values simply because the resonances in question are far enough apart with very narrow widths so the interference patterns between resonances cannot be seen.  It would, nevertheless, change the scattering angular distributions somewhat. However, the distributions are usually very close to isotropic at low energies so this too would have a small impact.

\squeezetable
\begin{table*}[htb]
    \caption{\label{table:reclassENDF} List of $^{52}$Cr resonances most reclassified in more than 50\% of the training events.
    The references corresponding to the lead authors below are given in Table \ref{table:52Cr_references}.
    $*$ Indicates that the given resonance's energy is above the upper limit of the resolved resonance region and that this resonance is present to provide a background contribution to the reconstructed cross section.
    %$**$ The 48.25253 keV $L$ assignment from Rohr et al. was incorrectly listed as the $J$ assignment in the EXFOR compilation.
    $L$ values in brackets indicate multiple possible assignments as per the original author.
    }
    \centering
    \begin{tabular}{@{}ccccccccccccccccccccc@{}}
        \toprule\toprule
              & Energy & Times  & Original & Times & Times & Times & \multicolumn{2}{c}{Agrawal} & Beer & Bilpuch & Bowman & \multicolumn{2}{c}{Brusegan} & \multicolumn{2}{c}{Carlton} & Hibdon & Kenny & \multicolumn{2}{c}{Rohr} & Stieglitz \\
        index & (keV) & reclassified & L & $L=0$ & $L=1$ & $L=2$ & $L$ & $J^\Pi$ & L & L & L & L & J & L & J & L & L & L & J & L \\
        \midrule
        358 & 1587.7*   & 951 & 0 &     & 111 & 840 &  &  &  &  &  &  &  &  &  &  &  &  &  &  \\
        311 & 1344.005* & 947 & 0 &     & 11  & 936 &  &  &  &  &  &  &  &  &  &  &  &  &  &  \\
        355 & 1497.4*   & 947 & 0 &     & 34  & 913 &  &  &  &  &  &  &  &  &  &  &  &  &  &  \\
        360 & 1730.9*   & 944 & 0 &     & 76  & 868 &  &  &  &  &  &  &  &  &  &  &  &  &  &  \\
        372 & 2260.9*   & 943 & 0 &     & 72  & 871 &  &  &  &  &  &  &  &  &  &  &  &  &  &  \\
        362 & 1832.6*   & 943 & 0 &     & 73  & 870 &  &  &  &  &  &  &  &  &  &  &  &  &  &  \\
        364 & 1888.5*   & 943 & 0 &     & 73  & 870 &  &  &  &  &  &  &  &  &  &  &  &  &  &  \\
        373 & 2307.7*   & 943 & 0 &     & 72  & 871 &  &  &  &  &  &  &  &  &  &  &  &  &  &  \\
        366 & 1959.5*   & 943 & 0 &     & 73  & 870 &  &  &  &  &  &  &  &  &  &  &  &  &  &  \\
        367 & 1999*     & 943 & 0 &     & 72  & 871 &  &  &  &  &  &  &  &  &  &  &  &  &  &  \\
        369 & 2178*     & 943 & 0 &     & 72  & 871 &  &  &  &  &  &  &  &  &  &  &  &  &  &  \\
        371 & 2238.2*   & 943 & 0 &     & 72  & 871 &  &  &  &  &  &  &  &  &  &  &  &  &  &  \\
        361 & 1791.9*   & 934 & 0 &     & 75  & 859 &  &  &  &  &  &  &  &  &  &  &  &  &  &  \\
        370 & 2204.2*   & 934 & 0 &     & 71  & 863 &  &  &  &  &  &  &  &  &  &  &  &  &  &  \\
        368 & 2078.5*   & 933 & 0 &     & 72  & 861 &  &  &  &  &  &  &  &  &  &  &  &  &  &  \\
        365 & 1929.2*   & 932 & 0 &     & 72  & 860 &  &  &  &  &  &  &  &  &  &  &  &  &  &  \\
        363 & 1868.2*   & 908 & 0 &     & 76  & 832 &  &  &  &  &  &  &  &  &  &  &  &  &  &  \\
        0 & 1.625867    & 870 & 1 & 178 &     & 692 & [1,2] & $\frac{3}{2}$ &  &  &  &  &  &  &  &  &  & 1 & $\frac{3}{2}$ &  \\
        2 & 22.95014    & 861 & 1 & 165 &     & 696 & [1,2] & $\frac{3}{2}$ & 1 &  &  & 1 & $\frac{3}{2}$ &  &  &  &  & 1 & $\frac{3}{2}$ &  \\
        4 & 27.59859    & 859 & 1 & 161 &     & 698 &  &  & 1 &  &  &  &  &  &  &  &  &  &  &  \\
        1 & 19.35777    & 854 & 1 & 165 &     & 689 &  &  &  &  &  &  &  &  &  &  &  &  &  &  \\
        3 & 24.85516    & 853 & 1 & 163 &     & 690 &  &  &  &  &  &  &  &  &  &  &  &  &  &  \\
        6 & 33.91804    & 848 & 1 & 158 &     & 690 &  &  & 1 &  &  &  &  &  &  &  &  &  &  &  \\
        7 & 34.32529    & 848 & 1 & 157 &     & 691 &  &  &  &  &  &  &  &  &  &  &  &  &  &  \\
        8 & 47.9453     & 827 & 1 & 134 &     & 693 &  &  &  &  &  &  &  &  &  &  &  &  &  &  \\
        9 & 48.25253    & 826 & 1 & 132 &     & 694 & 1 & $\frac{1}{2}^{-}$ & 1 &  &  & 1 &  & 1 & $\frac{3}{2}^{-}$ &  &  &  &  &  \\
        10 & 50.12      & 823 & 1 & 127 &     & 696 &  &  & 0 &  &  &  &  &  &  &  & 0 &  &  & 0 \\
        11 & 50.32564   & 822 & 0 &     & 228 & 594 & 0 & $\frac{1}{2}^{+}$ &  & 0 &  & 0 & $\frac{1}{2}$ & 0 & $\frac{1}{2}^{+}$ & 0 &  & 0 & $\frac{1}{2}$ &  \\
        342 & 1449      & 822 & 2 & 821 & 1   &     &  &  &  &  &  &  &  &  &  &  &  &  &  &  \\
        5 & 31.63829    & 815 & 0 &     & 184 & 631 & 0 & $\frac{1}{2}^{+}$ & 0 &  &  & 0 & $\frac{1}{2}$ &  &  &  &  & 0 & $\frac{1}{2}$ &  \\
        12 & 57.7435    & 801 & 1 & 104 &   & 697 & 1 & $\frac{1}{2}^{-}$ & 1 &  &  & 1 & $\frac{3}{2}$ & 1 & $\frac{1}{2}^{-}$ &  &  & 1 & $\frac{1}{2}$ &  \\
        359 & 1606.4*   & 757 & 0 &     & 79 & 678 &  &  &  &  &  &  &  &  &  &  &  &  &  &  \\
        13 & 68.23      & 740 & 1 & 68  &   & 672 &  &  &  &  &  &  &  &  &  &  &  &  &  &  \\
        50 & 258.1997   & 732 & 1 & 2   &   & 730 & [1,2] & $\frac{3}{2}$ & 1 &  &  & 1 & $\frac{3}{2}$ & 1 & $\frac{3}{2}^{-}$ &  &  & 1 & $\frac{3}{2}$ &  \\
        14 & 78.82184   & 724 & 1 & 34  &   & 690 &  &  & 1 &  &  &  &  &  &  &  &  &  &  &  \\
        46 & 247.3901   & 689 & 1 & 78  &   & 611 & [1,2] & $\frac{3}{2}$ &  &  &  & 1 & $\frac{3}{2}$ & 1 & $\frac{3}{2}^{-}$ &  &  & 1 & $\frac{3}{2}$ &  \\
        82 & 399.5334   & 575 & 1 & 0   &   & 575 & [1,2] & $\frac{3}{2}$ &  &  &  & 0 & $\frac{1}{2}$ & 1 & $\frac{1}{2}^{-}$ &  &  & 1 & $\frac{1}{2}$ &  \\
        49 & 251.6374   & 552 & 1 & 15  &   & 537 & [1,2] & $\frac{3}{2}$ &  &  &  & 1 & $\frac{1}{2}$ & 0 & $\frac{1}{2}^{+}$ &  &  & 1 & $\frac{1}{2}$ &  \\
        55 & 283.2405   & 551 & 1 & 99  &   & 452 & [1,2] & $\frac{3}{2}$ & 1 &  & 0 & 1 & $\frac{1}{2}$ & 1 & $\frac{1}{2}^{-}$ &  & 0 & 1 & $\frac{1}{2}$ &  \\
        22 & 121.94     & 542 & 0 &     & 60 & 482 & 0 & $\frac{1}{2}^{+}$ & 0 &  & 0 & 0 & $\frac{1}{2}$ & 0 & $\frac{1}{2}^{+}$ & 0 & 0 & 0 & $\frac{1}{2}$ &  \\
        52 & 265.1415   & 541 & 0 &     & 220 & 321 & 0 & $\frac{1}{2}^{+}$ &  &  &  & 0 & $\frac{1}{2}$ & 0 & $\frac{1}{2}^{+}$ &  &  & 0 & $\frac{1}{2}$ &  \\
        51 & 260.8998   & 525 & 1 & 4   &   & 521 &  &  &  &  &  &  &  &  &  &  &  &  &  &  \\
        48 & 250.5283   & 504 & 1 & 17  &   & 487 & [1,2] & $\frac{3}{2}$ &  &  &  & 1 & $\frac{1}{2}$ & 1 & $\frac{1}{2}^{-}$ &  &  & 1 & $\frac{1}{2}$ &  \\
        354 & 1493*     & 502 & 1 & 1   &   & 501 &  &  &  &  &  &  &  &  &  &  &  &  &  &  \\
       \bottomrule\bottomrule
    \end{tabular}
\end{table*}

\begin{table*}[htb]
  \caption{\label{table:52Cr_references} List of \nuc{52}{Cr} datasets in literature, as found in the EXFOR database~\cite{EXFOR-2,EXFOR},  in which partial complete spingroup assignments are given.  These do not necessarily agree with the choices of the ENDF/B-VIII.0 evaluators in Ref.~\cite{ND2010_Cr}.}
\centering
\begin{tabular}{llm{4in}}
\toprule\toprule
Reference                                   & EXFOR Entry & Spingroup determination method\\ \midrule
Agrawal et al. (1984) \cite{Agrawal:1984}   & 12830       & Shape analysis of transmission data coupled with measurement of the scattered neutron angular distribution. The forward/backward asymmetry allowed a determination of $\Pi$ and, coupled with $L$, a $J$ assignment.\\ \midrule
Allen et al. (1975) \cite{Allen:1975}       & 30393       & Used methods from Refs.~\cite{Steiglitz:1970,Musgrove:1974}.  Combined transmission and capture measurements at ORELA.  In some cases, the authors could measure capture width (mainly s-wave resonances) while in others only capture area ($L>0$).  $L$ assignments are based on the transmission measurement. While paper suggests this assignment was performed, only average resonance parameters are in the publication. 
\\ \midrule
Beer et al. (1975) \cite{Beer:1975}         & 20374       & Used methods from Refs.~\cite{Steiglitz:1970,Musgrove:1974}.  Combined transmission and capture measurements at Karlsruhe.  In some cases, the authors could measure capture width (mainly s-wave resonances) while in others only capture area ($L>0$).  $L$ assignments are based on the transmission measurement. \\ \hline
Bilpuch et al. (1961) \cite{Bilpuch:1961}   & 11599       & Shape analysis of transmission data.  \\ \midrule
Bowman et al. (1962) \cite{Bowman:1962}     & 11600       & Shape analysis of transmission data.  \\ \midrule
Brusegan et al. (1986) \cite{Brusegan:1986} & 22041       & Used methods from Refs.~\cite{Steiglitz:1970,Musgrove:1974}.  Combined transmission and capture measurements at GELINA.  In some cases, the authors could measure capture width (mainly s-wave resonances) while in others only capture area ($L>0$).  $L$ assignments are based on the transmission measurement. \\ \midrule
Carlton et al. (2000) \cite{Carlton:2000}   & 13840       & Shape analysis of transmission data coupled with measurement of the scattered neutron angular distribution. The forward/backward asymmetry allowed a determination of $\Pi$ and, coupled with $L$, a $J$ assignment.\\ \midrule
Hibdon (1957) \cite{Hibdon:1957}            & 11674       & Shape analysis of transmission data. \\ \midrule
Kenny et al. (1977) \cite{Kenny:1977}       & 30393       & Used methods from Refs.~\cite{Steiglitz:1970,Musgrove:1974}.  Combined transmission and capture measurements at ORELA.  In some cases, the authors could measure capture width (mainly s-wave resonances) while in others only capture area ($L>0$).  $L$ assignments are based on the transmission measurement. \\ \midrule
Rohr et al. (1989) \cite{Rohr:1989}         & 22131       & Shape analysis of transmission data. \\ \midrule
Stieglitz et al. (1970) \cite{Steiglitz:1970} & 10074 & Combined transmission measurements and capture measurements at RPI Linac.  The capture measurement registered capture events using a scintillator.  In some cases, the authors could measure capture width (mainly s-wave resonances) while in others only capture area ($L>0$).  $L$ assignments are based on the transmission measurement. \\ 
\bottomrule\bottomrule
\end{tabular}
\end{table*}

%===================================================================================================================
\section{Summary and conclusions}
\label{sec:conclusion}

In this paper, we have outlined the first application of machine learning to the long-standing problem of classifying neutron resonances by their appropriate quantum numbers.  We have outlined how we map statistical properties of resonances into OOD tests and then into features that can be used for resonance classification.   We have demonstrated the efficacy of our approach both with synthetic data and with a real study of the $^{52}$Cr ENDF/B-VIII.0 evaluation.  We noted problems with the use of capture widths when confronting older datasets.  

It is clear that our approach has many avenues for improvement:  
\begin{itemize}
\item There are many other features we wish to exploit including
\begin{inparaenum}[a)]
\item Dyson-Mehta $\Delta_3$ statistic and associated distribution,
\item use of the full spacing-spacing correlation, 
\item better capture width distributions, and
 \item per-resonance metadata, such as how were the quantum numbers determined and how confident are we in the determination.
\end{inparaenum}
%(a) Dyson-Mehta $\Delta_3$ statistic and associated distribution, (b) use of the full spacing-spacing correlation, (c) better capture width distributions, and (d) per-resonance metadata such as how were the quantum numbers determined and how confident are we in the determination. 
Some methods provide quite robust quantum number assignments while others only work well only for S-wave resonances.
\item We would like to continue testing the method, especially against experimental data where the full spingroup assignment is believed to be correct (e.g., polarized neutron and target experiments on actinides or from TRIPLE collaboration).
\item We would like to refine our classification strategies, including 
\begin{inparaenum}[a)]
	\item adopting iteration, namely refitting all OODs after each round of classification since Fig.~\ref{fig:validation_accuracy} demonstrates convergence under certain conditions;
	\item  adopting a staged approach where we first determine L, then move to full spingroup determination;
	\item optimizing choice of classifier and corresponding hyperparametrization;
	\item training and validating in sections of real resonance sequence data that are well-constrained experimentally; 
	\item exploring transfer learning to determine to what extent we can train on one nucleus's data and apply the classifier to another; and
	\item benchmark the quality of the classifier by incorporating additional performance metrics (such as precision, recall, ROC curves, etc.) in the analysis, better determining improvement routes.
\end{inparaenum}
\item In connection with the previous bullet, we would like to explore different measures of classification accuracy.  In this work, we used total accuracy.  As there are different numbers of resonances in each class (whether classifying by $L$ or spingroup), we have imbalanced sets of data.  In such this case, a balanced accuracy metric may be more appropriate \cite{balanced_accuracy}.
\item We would like to start a much broader discussion of the development of reproducible Uncertainty Quantification methods.  Such methods must address sensitivities to hyperparameters, feature weight, reclassification frequency, etc., to both the results of our classification and to the reconstructed neutron integrated and differential cross sections with the chosen spingroup assignments.
%(a) adopting iteration, namely refitting all OOD distributions after each round of classification since Fig.~\ref{fig:validation_accuracy} demonstrates convergence under certain conditions, (b) adopting a staged approach where we first determine L, then move to full spingroup determination, and (c) exploring transfer learning to determine to what extent we can train on one nucleus's data and apply the classifier to another.
\end{itemize}

In addition to these improvements, there are many other issues we must consider.  We have not attempted reclassification of a target nuclei with ground state $I^\Pi \ne 0^+$.  Therefore, we were able to ignore the $S$ quantum number and parity for the most part.  We also have not attempted to use fission resonances.  Also, there are questions about how doorway states and intermediate states might impact neutron width distributions.  Finally, we would like to understand what experimental effects may impact our results including, but not limited to, resonance sequence contamination from other isotopes and missing resonances.

%===================================================================================================================
\appendix
\section{Glossary of Machine-Learning terms}
\label{app:ML_Glossary}

To assist the reader who may not be fully familiarized with some of the common terms and expressions employed in Machine-Learning (ML) works, we briefly summarize some of the definitions as commonly adopted and/or as used in the current work:
\begin{itemize}
\item \textbf{Features:} set of relevant quantities used to describe and characterize the data points associated with the ML problem. Features can be vectorized and define a feature space that is assumed to represent well the input data.
\item \textbf{Labels:} Quantities associated with the output of a ML process. In other words, what the ML algorithm is attempting to predict. If labels are discrete quantities or objects or concepts, the ML algorithm is said to be a classifier.
\item \textbf{Training dataset:} Collection of data points of known labels that are used to train the ML algorithm. A trained algorithm is tuned to optimize the identification of labels from the training dataset.
\item \textbf{Testing dataset:} Collection of data points of known labels of similar origin as of the training set but that are not used in training. Their purpose is to assess how well the ML algorithm was trained to recognize data points similar to the training data set.
\item \textbf{Validation dataset:} Collection of data points of known labels that are compatible but independent (not of the  same origin)  of the training dataset. Their purpose is to assess how well the trained algorithm can perform in data points that it has never encountered before.
\item \textbf{Hyperparameters:} Parameters of the ML algorithm that can not be fully constrained by the model, and may be tuned to optimize the performance of the ML algorithm.
\item \textbf{Training seed:} The training subset randomly obtained after the input training data is randomly split in the classifier training process into a training and testing data set.
\item \textbf{Training event:} The definition of a trained classifier using a particular training seed. Because each training seed is a different sample of the complete training data, each training event will lead to a different classifier, and thus a different set of predictions.
\end{itemize}

%
%\fixme{In case we want to bore the reviewers with a lot of details about something.}

%\begin{theacknowledgments}
\section*{Acknowledgments}
%\end{theacknowledgments}
The authors would like to thank fruitful conversations with Declan Mulhall, Denise Neudecker, Michael Grosskopf and Vladimir Sobes.  We also would like to thank
Said Mughabghab for planting the seeds of this work with his decades long dedication to the development of the \textit{Atlas of Neutron Resonances}.

This work was supported by the Nuclear Criticality Safety Program, funded and managed by the National Nuclear Security Administration for the U.S. Department of Energy. Additionally, work at Brookhaven National Laboratory was sponsored by the Office of Nuclear Physics, Office of Science of the U.S. Department of Energy under Contract No. DE-SC0012704 with Brookhaven Science Associates, LLC. This project was supported in part by the Brookhaven National Laboratory (BNL), National Nuclear Data Center under the BNL Supplemental Undergraduate Research Program (SURP) and by the U.S. Department of Energy, Office of Science, Office of Workforce Development for Teachers and Scientists (WDTS) under the Science Undergraduate Laboratory Internships Program (SULI).

%===================================================================================================================
% BIBLIOGRAPHY
%===================================================================================================================

%\bibliographystyle{plainnat}  %% nome-ano
%\bibliographystyle{unsrt}    %% numero
\bibliographystyle{apsrev4-1} % temporarily commented out by fsd
\bibliography{brr_features}

\end{document}